\documentclass[aps,prl,twocolumn,superscriptaddress]{revtex4-2}
\usepackage{cancel}
\usepackage[normalem]{ulem}
\usepackage{bm}
\usepackage{amsmath}
\usepackage{amsthm}
\usepackage{natbib}
\usepackage{amssymb}
\usepackage{graphicx}
\usepackage{esint}

\makeatletter
\usepackage{times}
\usepackage{comment}
\usepackage{graphicx}
\usepackage{feynmf}
\usepackage{tabularx}
\usepackage{amsmath}
\usepackage{amstext}
\usepackage{amssymb}
\usepackage{xfrac}
\usepackage[colorlinks,citecolor=blue]{hyperref}
\usepackage{graphicx}
\usepackage{amsmath}
\usepackage{amstext}
\usepackage{amssymb}
\usepackage{amsfonts}
\usepackage{longtable,booktabs}
\usepackage{hyperref}
\usepackage{url}
\usepackage{subfigure}
\usepackage{dsfont}
\usepackage{booktabs}
\usepackage{amsbsy}
\usepackage{dcolumn}
\usepackage{amsthm}
\usepackage{bm}
\usepackage{esint}
\usepackage{multirow}
\usepackage{hyperref}
\usepackage{cleveref}
\usepackage{mathrsfs}
\usepackage{amsfonts}
\usepackage{amsbsy}
\usepackage{dcolumn}
\usepackage{bm}
\usepackage{multirow}
\usepackage{color}
\usepackage{extarrows}
\usepackage{datetime}
\usepackage[super]{nth}
\hypersetup{
	colorlinks=true,
	linkcolor=blue,
	filecolor=magenta,
	urlcolor=magenta,
}

\newtheorem{theorem}{Theorem}

\newtheorem{lemma}{Lemma}

\begin{document}
\title{Emergent Causality of Macroscopic Transport and Equilibration in Long-Range Quantum Systems}
\author{Hongchao Li}
\affiliation{Department of Physics, The University of Tokyo, 7-3-1 Hongo, Tokyo 113-0033, Japan}
\email{lhc@cat.phys.s.u-tokyo.ac.jp}

\author{Marius Lemm}
\affiliation{Department of Mathematics, University of T\"ubingen, 72076 T\"ubingen, Germany}
\email{marius.lemm@uni-tuebingen.de}

\date{\today}
\begin{abstract}
Long-range couplings in quantum many-body systems challenge conventional notions of locality and causality. Here, we show that macroscopic particle transport nevertheless exhibits emergent causality --- an effective light cone that becomes strict in the thermodynamic limit. For power-law hopping with a decay exponent $\alpha\in(D,D+1)$ in $D$ spatial dimensions, the emergent causality cone expands super-ballistically, and we establish that its scaling is optimal. Our results imply lower bounds on the equilibration time of long-range systems. Finally, we construct an example of a long-range free-fermion chain with $\alpha\in (1,3/2)$, whose macroscopic density equilibrates super-ballistically, i.e., in a time sub-linear in the system size. While long-range interactions are usually associated with slow equilibration due to increased energy barriers, our example demonstrates that long-range hopping can lead to accelerated equilibration.
\end{abstract}
\maketitle
\emph{Introduction.---}A central pillar of relativity is the principle of causality, manifesting as the light cone, which sets a fundamental speed limit on information propagation. Remarkably, an analogous effective light cone arises for information propagation in short-range quantum spin systems on lattices, as established by the Lieb–Robinson (LR) bound \cite{Lieb1972}. The LR bound thus provides an approximate notion of causality in nonrelativistic quantum many-body systems, which has been extensively exploited in quantum information theory and condensed-matter physics. In particular, the LR bound is the decisive principle that provides, e.g., clustering of correlations \cite{NachtergaeleSims2006,Hastings2006long} and the area law of entanglement entropy in gapped systems~\cite{Hastings_2007}, the stability of topological phases~\cite{Hastings2005adiabatic,Bravyi2006,Hastings2004LSM}, and the simulatability of quantum many-body systems~\cite{Osborne2006,haah2021quantum,yu2025optimization,Tran2019}. The LR bound has also been extended to various other systems, including long-range spin systems and fermionic systems~\cite{Tran_hierarchy_2020,Kuwahara_light2020,KuwaharaPRL2021,Tran2019,ChenLucas:2019,Colmenarez2020,teufel2026lossless}, general interacting bosonic systems~\cite{Kuwahara2021,Yin2022,Faupin_LR2022,Lemm2023info,Kuwahara2024,elgart2026lieb,KuwaharaLemm2026,Carlalemm2026liebrobinsonbounds,rubiliani2026lieb}, open quantum systems~\cite{LRopen_2010,Barthel2012,XHan2018,Sweke_2019,Li2024BEC,Rahul2024,lemm2026dissipation,rubiliani2026lieb_open}, and, more recently, systems in continuous space~\cite{lemm2026lieb_continuum,Hinrichs2025,Bachmann2025,gebert2020lieb}.

Despite the remarkably broad applicability of the LR bound, it remains insufficient to characterize macroscopic particle transport and equilibration, as it substantially underestimates the suppression of collective particle motion \cite{Faupin2022}. This limitation becomes particularly pronounced in long-range systems, where algebraically decaying hopping allows faster propagation and gives rise to algebraic tails outside conventional light cones~\cite{Vu_2024,Hongchao2026,lemm2023microscopic}. Recently, a stronger form of locality was established for short-range systems by introducing the new dynamical principle of macroscopic suppression of supersonic macroscopic transport (MASSMAT)~\cite{Faupin2025}. The principle shows that the probability of transporting a finite fraction $\Delta\theta\in(0,1]$ of all $N$ particles beyond a ballistic light cone is exponentially suppressed in $\Delta\theta N$. Consequently, the light cone becomes effectively strict for macroscopic transport in the thermodynamic limit, revealing an unforeseen sharp emergent causality for short-range many-body systems. Meanwhile, long-range couplings can profoundly affect information propagation \cite{Tran_hierarchy_2020,Kuwahara_light2020,KuwaharaPRL2021,Tran2019,ChenLucas:2019,Colmenarez2020} and equilibration behavior \cite{kastner2011diverging,Neyenhuis2017Prethermalization,MonthusGarel2013DysonBarriers,Lenrose2019,liu2019confined}. These observations raise two fundamental questions: Does the emergent strict causality principle uncovered in Ref. \cite{Faupin2025} persist in long-range systems, and what are the consequences for equilibration?

\begin{figure}
	\IfFileExists{macroscopic_suppression.png}{%
    \includegraphics[width=0.98\columnwidth]{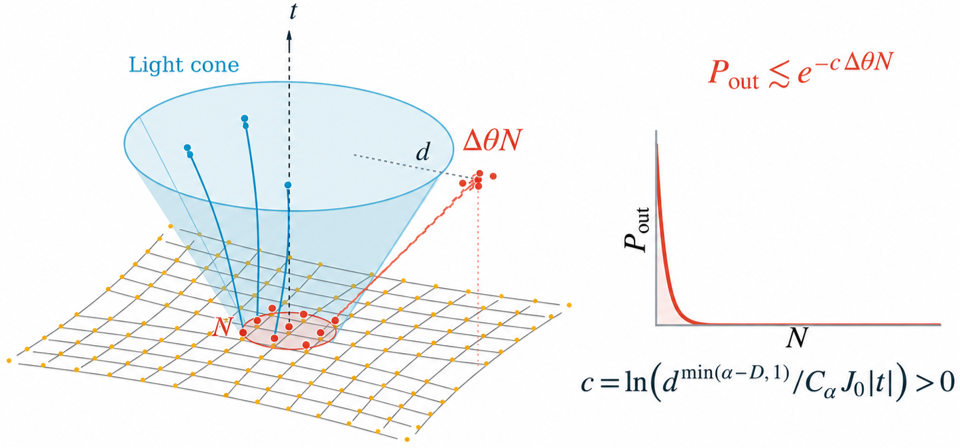}%
    }{\fbox{\parbox{0.9\columnwidth}{\centering Original figure file\\\texttt{macroscopic\_suppression.png}\\not supplied with the draft.}}}
	\caption{Schematics of the suppression of macroscopic particle transport in long-range quantum many-body bosonic systems. There exists a light-cone structure $d^{\min(\alpha-D,1)}=C_{\alpha}J_0|t|$ (with $d$ replaced by $d/\ln(2+d)$ at $\alpha=D+1$), where $\alpha$ denotes the decay exponent of long-range hopping, $J_0$ denotes the hopping amplitude, and $C_{\alpha}$ is an $O(1)$ constant. The probability of transporting $\Delta\theta N$ particles outside the light cone exponentially decreases with $\Delta\theta N$, indicating that a macroscopic transport violating causality is rigorously prohibited in the thermodynamic limit even in long-range systems.}
	\label{fig:transport}
\end{figure}

In this Letter, we answer both of these questions. First, we establish emergent causality for macroscopic particle transport in quantum systems with power-law hopping and arbitrary density-dependent interactions. Unlike previous bounds on long-range macroscopic transport~\cite{Vu_2024,Hongchao2026,lemm2023microscopic}, our results show that the probability of transporting a macroscopic number of particles beyond the effective light cone is exponentially suppressed in the number of particles transported, as illustrated in Fig.~\ref{fig:transport}. For a decay exponent $\alpha>D+1$ in $D$ spatial dimensions, the effective light cone remains ballistic. Remarkably, for $D<\alpha<D+1$, it expands super-ballistically, with distance scaling as $d\sim t^{1/(\alpha-D)}\gg t$. We demonstrate that these scalings are optimal through explicit state-transfer protocols.

Furthermore, we apply these causal constraints to derive lower bounds on equilibration time whenever equilibration requires macroscopic particle redistribution. We further demonstrate that long-range hopping can accelerate equilibration by constructing a long-range free-fermion chain with $1<\alpha<3/2$, in which the macroscopic density of any initial state equilibrates super-ballistically, namely in time $t_{\mathrm{eq}}\sim L^{\alpha-1/2}\ll L$. While long-range interactions can slow relaxation by enhancing energy barriers in certain Ising-type systems compared with their short-range counterparts~\cite{kastner2011diverging,Neyenhuis2017Prethermalization,MonthusGarel2013DysonBarriers,Lenrose2019,liu2019confined}, our results demonstrate that long-range hopping can instead accelerate macroscopic equilibration to a super-ballistic timescale. 

\emph{The model.---}We consider a quantum many-body system on a finite $D$-dimensional lattice composed of bosons or fermions, where $\Lambda$ denotes all the sites. We define $Z\subseteq\Lambda$ as a subset of the lattice sites on $\Lambda$, and the number of sites in $Z$ is denoted by $|Z|$. We assume that the lattice satisfies the uniform volume-growth bound:
\begin{equation}\label{eq: volume_grow}
    \sup_{i\in\Lambda}|i[r]\backslash i[r-1]|\leq \varphi r^{D-1},\qquad r\in\mathbb N,
\end{equation}
where $i[r]:=\{j\in\Lambda:|i-j|\leq r\}$, $A\backslash B$ represents the set of sites in the set $A$ that are not included in set $B$, and $\varphi$ is an $O(1)$ constant determined by the lattice structure. We consider the Hamiltonian
\begin{equation}\label{eq: Hamiltonian}
    H(t) := \sum_{i \neq j \in \Lambda} J_{i j}(t) a_i^{\dagger} a_j + \sum_{Z \subseteq \Lambda} h_Z,
\end{equation}
where ${a_i}$ and $a_i^{\dagger}$ are the annihilation and creation operators on the site $i \in \Lambda$.  The long-range hopping amplitude $J_{ij}(t)$ is locally integrable in $t$ and is assumed to be Hermitian: $J_{ij}=J_{ji}^{*}$, and obeys a power-law upper bound $\left| {{J_{ij}}(t)} \right| \le J_0/[1+|i-j|]^{\alpha}$ for a positive constant $J_0$, where $|i-j|$ represents the Euclidean distance (or periodic Euclidean distance on a torus) between the site $i$ and the site $j$ on the lattice, and the power-law decay exponent $\alpha$ satisfies $\alpha>D$ to ensure summability of the hopping amplitudes. The term $h_Z$ is an arbitrary self-adjoint function of local particle-number operators $\{n_i\}_{i\in Z}$ where $n_i:=a_i^{\dagger}a_i$. This class includes both short- and long-range density-dependent interactions, such as the on-site Bose–Hubbard interaction $h_Z=Un_i^2$, where $U$ is the interaction strength, and the subset $Z$ is taken to be the single site $i$. Our setup also encompasses spinful models and thus includes the Fermi-Hubbard Hamiltonian as well.  
 
We will study the unitary time evolution  generated by $H(t)$ by solving the time-dependent Schrödinger equation $\mathrm{i}  \frac{d}{dt}|\psi_t\rangle=H(t)|\psi_t\rangle$
with $\hbar=1$. For simplicity, we focus on time-independent Hamiltonians in the following, for which the unitary evolution is given by $U(t,0)=e^{-\mathrm{i}Ht}$. Nevertheless, the proof carries over verbatim to time-dependent Hamiltonians~\footnote{For time-dependent Hamiltonians, the unitary time evolution is governed by the time-ordered exponential $\mathcal{T}\exp[-\mathrm{i}\int_0^tH(s)\,ds]$. The proof is exactly the same because all we use is that the upper bound on the hopping coefficients $|J_{ij}(t)|$ is time-independent.}. 

\emph{The main result.---}
In this work, we investigate the transport of a macroscopic number of particles from a region $Y$ to a distant region $X$ on the lattice $\Lambda$, separated by a distance denoted as $d_{XY}:=\min_{x\in X,y\in Y}|x-y|$. Since the Hamiltonian \eqref{eq: Hamiltonian} has the global U(1) symmetry, the total particle number is conserved in the dynamics. Given the total particle number $N$, we derive an upper bound on the probability of transporting $\Delta\theta N$ particles from the region $Y$ to the region $X$ within time $t$, where $\Delta\theta\in(0,1]$ represents the ratio of the transported particles. We use $\theta_{\mathrm i}$ for the allowed initial fraction outside $Y$ and $\theta_{\mathrm f}$ for the required final fraction inside $X$. The macroscopic transport can be captured by the spectral projection operator $P_{\bar{N}_X\geq\theta_{\mathrm f}}$, where $N_X:=\sum_{i\in X}n_i$ and $\bar{N}_X:=N_X/N$ denotes the fraction of particles in the region $X$.  If the normalized initial state satisfies the condition $|\psi_0\rangle=P_{\bar{N}_Y\geq1-\theta_{\mathrm i}}|\psi_0\rangle$, the probability of transporting at least a fraction $\Delta\theta:=\theta_{\mathrm f}-\theta_{\mathrm i}$ of all the particles within the time $t$ can be expressed as
\begin{equation}
    \langle\psi_t|P_{\bar{N}_X\geq\theta_{\mathrm f}}|\psi_t\rangle=\|P_{\bar{N}_X\geq\theta_{\mathrm f}}|\psi_t\rangle\|^2.
\end{equation}

Our main result is summarized in the following theorem.
\begin{theorem}\label{th:main}
  Let $X,Y\subseteq\Lambda$ be two subsets separated by a distance $d_{XY}=d\geq1$. For the lattice $\Lambda$ satisfying Eq. \eqref{eq: volume_grow}, the Hamiltonian \eqref{eq: Hamiltonian} and any $0\leq\theta_{\mathrm i}<\theta_{\mathrm f}\leq1$, with $\Delta\theta:=\theta_{\mathrm f}-\theta_{\mathrm i}\in(0,1]$, there always exists a constant $C_{\alpha}>0$ depending only on the lattice geometry, $\alpha$, and $\Delta\theta$, such that
  \begin{equation}\label{eq: bound_1}
      \|P_{\bar{N}_X\geq\theta_{\mathrm f}}U(t,0)P_{\bar{N}_Y\geq1-\theta_{\mathrm i}}\|\leq\left(\frac{C_{\alpha}J_0|t|}{d^{\min(\alpha-D,1)}}\right)^{\Delta\theta N/2}
  \end{equation}
  in the large-system limit: $|\Lambda|,d\to\infty$, where $\|O\|:=\max_{\||\psi\rangle\|=1}\|O|\psi\rangle\|$ is the norm of the operator $O$. At $\alpha=D+1$, replace the denominator $d$ in~\eqref{eq: bound_1} by $d/\ln(2+d)$.
\end{theorem}
The theorem is proved in Supplemental Material~\cite{SupplementaryMaterial}. We note that this result rigorously holds in the limit $d\to\infty$ without replacing the distance $d$ in Eq. \eqref{eq: bound_1} with $d/\ln(2+d)$ for $\alpha=D+1$. With Eq. \eqref{eq: bound_1}, we can estimate the upper bound for the quantity $\langle\psi_t|P_{\bar{N}_X\geq\theta_{\mathrm f}}|\psi_t\rangle$ since
\begin{align}
    \langle\psi_t|P_{\bar{N}_X\geq\theta_{\mathrm f}}|\psi_t\rangle&=\|P_{\bar{N}_X\geq\theta_{\mathrm f}}U(t,0)|\psi_0\rangle\|^2\nonumber\\
    &\leq\|P_{\bar{N}_X\geq\theta_{\mathrm f}}U(t,0)P_{\bar{N}_Y\geq1-\theta_{\mathrm i}}\|^2\nonumber\\
    &\leq\left(\frac{C_{\alpha}J_0|t|}{d^{\min(\alpha-D,1)}}\right)^{\Delta\theta N},
\end{align}
with the same replacement at $\alpha=D+1$. This serves as the upper bound for the probability of transporting $\Delta\theta N$ particles from the region $Y$ to the region $X$. This result implies an effective power-law light cone: $t\propto d^{\min(\alpha-D,1)}$ for $\alpha\ne D+1$, and $t\propto d/\ln(2+d)$ at $\alpha=D+1$. For $D<\alpha<D+1$, this super-ballistic scaling can be heuristically understood from a translationally invariant, non-interacting model with isotropic hopping $J(\bm r)\sim J_0/|\bm r|^\alpha$, whose Fourier transformation gives $|\omega(k)-\omega(0)|\sim J_0 k^{\alpha-D}$ at small $k=|\bm k|$. The group velocity therefore scales as $|v_g(k)|=|d\omega(k)/dk|\sim J_0 k^{\alpha-D-1}$, and modes with $k\sim d^{-1}$ have a propagation timescale $t\sim d/|v_g(d^{-1})|\sim d^{\alpha-D}/J_0$, consistent with the light-cone exponent above~\cite{Defenu_RMP_2023}. When $\alpha>D+1$, the light-cone structure is fixed as a linear one $t\propto d$, which does not change with the decay exponent $\alpha$. Compared with the macroscopic transport limits in long-range systems~\cite{Vu_2024}, there exists an additional power $\Delta\theta N$ which exponentially suppresses the macroscopic transport of particles outside the light cone. Our results show the emergent thermodynamic causality, i.e., the violation of causality is rigorously prohibited in the thermodynamic limit: $N\to\infty$. We also emphasize that this result holds for arbitrary hopping amplitudes and spatial dimensions satisfying the stated assumptions, and independently of the strength and range of density-dependent interactions, showing that such interactions cannot overcome the universal constraints on macroscopic particle transport.


Next, we establish the optimality of the power-law light-cone scaling in Eq. \eqref{eq: bound_1} for $\alpha\ne D+1$, using an explicit state-transfer protocol on a sufficiently large cubic lattice. Consider the initial state $|\psi_y^N\rangle=(a_y^{\dagger})^N|\text{vac}\rangle/\sqrt{N!}$, in which all $N$ bosons occupy the site $y$. Adapting the state transfer protocol in Ref. \cite{Tran_hierarchy_2020}, we construct a number-preserving Hamiltonian $H(t)$ satisfying the prescribed hopping constraints, such that 
\begin{equation}
    \langle\psi_x^N|U(t,0)|\psi_y^N\rangle=1,
\end{equation}
where $|\psi_x^N\rangle:=(a_x^{\dagger})^N|\text{vac}\rangle/\sqrt{N!}$ represents the state with all the bosons transported to the site $x$, and the time $t$ for the state transfer scales as 
\begin{equation}
t\propto
\begin{cases}
|x-y|^{\alpha-D}/J_0, & D<\alpha<D+1,\\
|x-y|/J_0, & \alpha\geq D+1,
\end{cases}
\end{equation}
with constants independent of $N$ and $|x-y|$. The specific form of the Hamiltonian and the proof is provided in Supplemental Material~\cite{SupplementaryMaterial}. At $\alpha=D+1$, the protocol leaves a logarithmic gap. The extension to $N$ bosons follows from the quadratic structure of the Hamiltonian, which linearly transforms the creation operators as $U(t,0)a_j^{\dagger}\bigl(U(t,0)\bigr)^{\dagger}=\sum_{i\in\Lambda}u_{ij}(t)a_i^{\dagger}$. Therefore, the overlap between the evolved state $U(t,0)|\psi_y^N\rangle$ and the target state $|\psi_x^N\rangle$ is given by
\begin{align}
    \langle\psi_x^N|U(t,0)|\psi_y^N\rangle&=\frac{1}{N!}\langle\text{vac}|a_x^N\left(\sum_{i\in\Lambda}u_{iy}(t)a_i^{\dagger}\right)^N|\text{vac}\rangle\nonumber\\
    &=u_{xy}(t)^N.
\end{align}
Thus, a protocol achieving $u_{xy}(t)=1$ can also transfer all $N$ bosons perfectly in the same time as a single particle.

\emph{Application.---} We next apply Theorem \ref{th:main} to estimate the equilibration time of translationally invariant quantum many-body systems. We first decompose the Hilbert space $\mathcal{H}$ into the equilibrium subspace $\mathcal{H}_{\text{eq}}$ and the nonequilibrium subspace $\mathcal{H}_{\text{neq}}$. Fix a partition $\Lambda=\bigsqcup_{j=1}^m B_j$ with $m$ being independent of $|\Lambda|$ and $|B_j|\geq\eta|\Lambda|$ for some fixed $\eta>0$. The density of particles on the lattice is given by $\bar{n}=N/|\Lambda|$. Then by introducing the regional number operators $N_j:=N_{B_j}=\sum_{i\in B_j}n_i$ and $\Delta N_j:=N_j-\bar n|B_j|$, we can define the equilibrium subspace as~\cite{HaraKoike2026}
\begin{equation}
    \mathcal{H}_{\text{eq}}:=\text{span}\{|\psi_k\rangle;|\langle\psi_k|\Delta N_j|\psi_k\rangle|\leq\epsilon\bar{n}|B_j|\ \text{for all $j$}\},
\end{equation}
where $\{|\psi_k\rangle\}$ is the occupation-number basis of $\mathcal H$ and $0<\epsilon\ll1$ is fixed as a small constant. Set $\mathcal H_{\text{neq}}=\mathcal H_{\text{eq}}^\perp$. Thus, equilibration requires the measured densities in the macroscopic regions to be close to $\bar{n}$ up to an error $\epsilon\bar{n}$. Hence, the projection operator onto the equilibrium (nonequilibrium) subspace is $P_{\text{eq}}:=\sum_{j:\,|\psi_j\rangle\in\mathcal{H}_{\text{eq}}}|\psi_j\rangle\langle\psi_j|$ ($P_{\text{neq}}:=\sum_{j:\,|\psi_j\rangle\in\mathcal{H}_{\text{neq}}}|\psi_j\rangle\langle\psi_j|=1-P_{\text{eq}}$). Given a time $T$ and an initial state $|\Psi(0)\rangle$, we introduce the set of nonequilibrium times during the time interval $[0,T]$ as $\mathcal{T}^{[0,T]}_{\delta,|\Psi(0)\rangle}:=\{t\in[0,T];\langle\Psi(t)|P_{\text{neq}}|\Psi(t)\rangle\geq\delta\}$. We call a sequence $t_{\mathrm{eq}}=t_{\mathrm{eq}}(|\Lambda|)>0$ an equilibration timescale if, for $T\geq t_{\mathrm{eq}}(|\Lambda|)$ and every fixed $\delta\in(0,1)$,
\begin{equation}
    \frac{1}{T}\left|\mathcal{T}^{[0,T]}_{\delta,|\Psi(0)\rangle}\right|\rightarrow0,
\end{equation}
in the thermodynamic limit: $N,|\Lambda|\to\infty$.

We have the following result:
\begin{theorem}\label{th:equil-lower}
    Assume that the particle number density $N/|\Lambda|$ is a fixed positive constant. For a cubic lattice $\Lambda=(\mathbb{Z}/L\mathbb{Z})^D$ with side length $L$ under the periodic boundary condition and the Hamiltonian \eqref{eq: Hamiltonian}, the normalized initial state satisfies $N_Y|\Psi(0)\rangle=N|\Psi(0)\rangle$, where $Y:=\Lambda\cap\prod_{i=1}^D[0,L/a_i]$, with fixed $a_1,\ldots,a_D>1$. Assume that $d_{B_j,Y}\geq\eta L$ for some $j$, using the periodic distance. Every equilibration timescale $t_{\text{eq}}$ satisfies, for all sufficiently large $L$,
\begin{equation}\label{eq: time_eq}
    t_{\mathrm{eq}}\geq\frac c{J_0}
    \begin{cases}
    L^{\alpha-D}, & D<\alpha<D+1,\\
    L/\ln(L), & \alpha=D+1,\\
    L, & \alpha>D+1,
    \end{cases}
\end{equation}
where $c>0$ is independent of $L$ and of the initial state.
\end{theorem}
The detailed proof is provided in Supplemental Material~\cite{SupplementaryMaterial}. Such spatially confined initial states are widely used to study macroscopic equilibration. The same argument applies to a cubic lattice with half of the sites occupied by particles, i.e., the initial particles sit at $[0,L/2]\times[0,L]^{D-1}$. This is a natural implication from Theorem \ref{th:main}. The initial particles are all restricted in a box on the lattice. Equilibration can be achieved only if the measured particle density in a distant region reaches at least $(1-\epsilon)N/L^D$ with probability tending to one. Therefore, a macroscopic number of particles should be transported to the initially empty sites from the box within time $t$ to reach equilibration. However, due to the exponential suppression by $\Delta\theta N$ in Eq. \eqref{eq: bound_1}, such transportation is exponentially unlikely for sufficiently small fixed $J_0t/L^{\min(\alpha-D,1)}$, with the logarithmic correction at $\alpha=D+1$, giving the stated lower bound. The previous result on optimal equilibration time for finite-range-hopping fermions on the cubic lattices is given by $t= O(Lg(L))$, where $g(L)\to\infty$ as $L\to\infty$~\cite{HaraKoike2026}. In contrast to Ref.~\cite{HaraKoike2026}, which requires $\eta=O(1)$, our result allows $\eta=1/|\Lambda|$, thereby extending the analysis of equilibration to single-site resolution. Equilibration at any spatial scale cannot occur on timescales shorter than the lower bound established in Eq.~\eqref{eq: time_eq}.

 In addition, we emphasize that the lattice structure is not necessarily the cubic lattice and can also be generalized to the honeycomb lattice and the triangular lattice under the periodic boundary condition. The proof is essentially the same. The initial distribution can be further relaxed: it suffices that approaching equilibrium requires transporting a finite fraction of the particles over a distance proportional to $L$.

 In the regime $\alpha\in(D,D+1)$, Theorem \ref{th:equil-lower} raises the intriguing possibility of super-ballistically accelerated equilibration. To demonstrate this possibility, we construct an explicit example of a long-range non-interacting fermionic chain with $\alpha\in (1,3/2)$ and $D=1$, in which any initial state equilibrates on a characteristic timescale $t_{\text{eq}}\sim L^{\alpha-1/2}\ll L$, growing sub-linearly with the system size $L$. The details of the model and the proof of accelerated equilibration are provided in Supplemental Material~\cite{SupplementaryMaterial}. Our construction provides a surprisingly simple and transparent proof that long-range hopping can parametrically accelerate equilibration in (quasi-)particle models. The result is particularly striking because long-range interactions can instead \textit{slow down} equilibration in certain false-vacuum problems by increasing the energy barriers associated with domain walls~\cite{kastner2011diverging,Neyenhuis2017Prethermalization,MonthusGarel2013DysonBarriers,liu2019confined}. Thus, our findings reveal the contrasting effects of long-range couplings on equilibration, ranging from barrier-induced slowing to super-ballistically accelerated relaxation.

\emph{Skeleton of the Proof.---} Here we provide the skeleton of the proof of Theorem \ref{th:main}. We introduce the geometric exponential tilting based on the particle-number operator~\cite{Faupin2025,Sigallr2025}:
\begin{equation}
    T_{\mu}=\exp\left(\mu\sum_{x\in\Lambda}F_xn_x\right),
\end{equation}
where $F_x$ is a designed real-valued function. The function $F_x$ is required to satisfy $F_x=1$ for $x\in X$, $F_x=-1$ for $x\in Y$, and $|F_x-F_y|\leq2\min\{|x-y|/d,1\}$. Then the function gives a weight $1$ for the particles in the region $X$ and a weight $-1$ for the particles in the region $Y$. By inserting $T_{\mu}T_{\mu}^{-1}$ into the left-hand side of Eq. \eqref{eq: bound_1}, we obtain
\begin{align}
    &\|P_{\bar{N}_X\geq\theta_{\mathrm f}}e^{-\mathrm{i}Ht}P_{\bar{N}_Y\geq1-\theta_{\mathrm i}}\|\nonumber\\
    \leq&\|P_{\bar{N}_X\geq\theta_{\mathrm f}}T_{\mu}^{-1}\|\|T_{\mu}e^{-\mathrm{i}Ht}T_{\mu}^{-1}\|\|T_{\mu}P_{\bar{N}_Y\geq1-\theta_{\mathrm i}}\|.
\end{align}
The proof of Theorem \ref{th:main} is decomposed into two parts: the upper bounds of the terms $\|P_{\bar{N}_X\geq\theta_{\mathrm f}}T_{\mu}^{-1}\|$ and $\|T_{\mu}P_{\bar{N}_Y\geq1-\theta_{\mathrm i}}\|$, and the upper bound of the term $\|T_{\mu}e^{-\mathrm{i}Ht}T_{\mu}^{-1}\|$. For the first part, we can rewrite the term $\|P_{\bar{N}_X\geq\theta_{\mathrm f}}T_{\mu}^{-1}\|$ as
\begin{equation}
    \|P_{\bar{N}_X\geq\theta_{\mathrm f}}T_{\mu}^{-1}\|=\max_{\|\psi\|=1}\langle\psi|P_{\bar{N}_X\geq\theta_{\mathrm f}}\exp\left[-\mu\sum_{x\in\Lambda}F_xn_x\right]|\psi\rangle,
\end{equation}
which reaches the maximal value if $|\psi\rangle=P_{\bar{N}_X\geq\theta_{\mathrm f}}|\psi\rangle$. This condition imposes constraints on $T_{\mu}^{-1}$, which makes it be upper bounded by $\exp[-\mu N(\theta_{\mathrm f}-(1-\theta_{\mathrm f}))]=\exp[-\mu N(2\theta_{\mathrm f}-1)]$. Similarly, we can also upper bound the term $\|T_{\mu}P_{\bar{N}_Y\geq1-\theta_{\mathrm i}}\|$ by $\exp[\mu N(2\theta_{\mathrm i}-1)]$. Therefore, the two projector norms contribute a factor $\exp[-2\mu N(\theta_{\mathrm f}-\theta_{\mathrm i})]=\exp(-2\mu\Delta\theta N)$. By choosing
\begin{equation}
    \mu\simeq\ln(d^{\min(\alpha-D,1)}/(CJ_0|t|)),
\end{equation}
with $d$ replaced by $d/\ln(2+d)$ at $\alpha=D+1$ and the precise choice given in Supplemental Material~\cite{SupplementaryMaterial}, the first part leads to the structure of the bound on the right-hand side of Eq. \eqref{eq: bound_1}. For the second part, we show that the contribution from the evolution is bounded by $\exp[C'N]$, where $C'$ is independent of the distance $d$ and the time $t$. Using $T_{\mu}\exp(-\mathrm{i}tH)T_{\mu}^{-1}=\exp(-\mathrm{i}tT_{\mu}HT_{\mu}^{-1})$, we obtain the transformed Hamiltonian
\begin{equation}
    T_{\mu}HT_{\mu}^{-1}=\sum_{x,y\in\Lambda}J_{xy}e^{\mu(F_x-F_y)}a_x^{\dagger}a_y+\sum_{Z\subseteq\Lambda}h_Z.
\end{equation}
Since the density-dependent interactions commute with $T_\mu$, they do not contribute to $\operatorname{Im}(T_\mu HT_\mu^{-1})$, which controls the norm of the transformed evolution operator. By dividing the sum $\sum_y$ into two regimes: $|x-y|\leq d/2\mu$ and $|x-y|>d/2\mu$, and using the stated Lipschitz bound on $F_x$, we can show that $|t|\,\|\operatorname{Im}(T_{\mu}HT_{\mu}^{-1})\|$ is bounded by $C'N$, where $C'>1$ only depends on the lattice geometry and the decay exponent $\alpha$. Combining the contributions from the two parts, we have the bound in Eq. \eqref{eq: bound_1}. See the detailed proof in Ref.~\cite{SupplementaryMaterial}.

\emph{Possible Experimental Situation.---} Here we propose experimental protocols to test our theoretical predictions using quantum simulation with the platform of Rydberg neutral atoms, following the preparation and measurement scheme proposed in Ref.\cite{Hongchao2026}. The Rydberg atoms can be arranged in an optical lattice or on a tweezer array to simulate the lattice structure~\cite{bloch2005ultracold}. The boson number distribution can be encoded into the spin systems, which serve as hard-core bosonic systems. The long-range hopping and the interactions can be set and tuned effectively by an off-resonant laser field~\cite{Macri2014,Defenu_RMP_2023,Hollerith2022PRL,PRXQuantum.3.020303,wuster2011excitation,PhysRevResearch.6.033282}, with the effective Hamiltonian taking the form $H=\sum_{i,j}J_{ij}\sigma_i^+\sigma_j^-+\mathrm{h.c.}+\sum_{i,j}V_{ij}n_in_j$. Spatially selective excitation allows the preparation of an initial state with spins polarized up in the source region $Y$  (occupied) and down in the distant target region $X$ (vacuum). After an evolution for a time $t$, one can measure the transferred particle number by detecting the site-resolved distribution of the Rydberg atom states through fluorescence imaging~\cite{Labuhn2016,Scholl2023}. Repeating the experiment and reconstructing the distribution of particles, one can directly estimate the probability of transportation and observe the equilibration time. 

\emph{Conclusion.---}In this Letter, we have rigorously established macroscopic speed limits on particle transport in long-range quantum many-body systems. For long-range hopping, the probability of transporting a finite fraction of the particles beyond an effective light cone is exponentially suppressed in the number of transported particles. Consequently, the light cone becomes effectively strict in the thermodynamic limit, revealing emergent thermodynamic causality even in long-range systems. We apply the transport constraints to derive lower bounds on the equilibration time of systems initially confined to spatially localized regions. Notably, we also construct a long-range fermionic model in which macroscopic density equilibration is super-ballistically accelerated.

Our Letter suggests several directions for future research. One is to establish stronger bounds on information propagation in quantum spin systems. Conventional Lieb–Robinson bounds for long-range systems constrain operator spreading but retain algebraically decaying tails, allowing information leakage outside the light cone. We conjecture a stronger bound whose suppression exponent grows with the amount of transmitted information, quantified through the Holevo capacity, thereby strongly suppressing the transmission of macroscopic amounts of information outside the light cone. Another direction is to construct concrete Hamiltonians that reach the lower bound of the equilibration time in Eq.~\eqref{eq: time_eq} for long-range systems.

\emph{Acknowledgements.---} We are grateful to Xie-Hang Yu, Akihiro Hokkyo, Tomotaka Kuwahara, and Zongping Gong for fruitful discussions. H.~L. is supported by Forefront Physics and Mathematics Program to Drive Transformation (FoPM), a World-leading Innovative Graduate Study (WINGS) Program, the University of Tokyo. H.~L. also acknowledges JSPS KAKENHI (Grant No.~JP24KJ0824). The research of M.~L. is supported by the Deutsche Forschungsgemeinschaft (DFG, German Research Foundation)  through the grants TRR 352 – Project-ID 470903074 and FOR 5413 – Project-ID 465199066, as well as by the European Union through ERC Starting Grant MathQuantProp, Grant Agreement 101163620.\footnote{Views and opinions expressed are however those of the authors only and do not necessarily reflect those of the European Union or the European Research Council Executive Agency. Neither the European Union nor the granting authority can be held responsible for them.} 

\bibliography{MyNewCollection}

\begin{thebibliography}{65}%
\makeatletter
\providecommand \@ifxundefined [1]{%
 \@ifx{#1\undefined}
}%
\providecommand \@ifnum [1]{%
 \ifnum #1\expandafter \@firstoftwo
 \else \expandafter \@secondoftwo
 \fi
}%
\providecommand \@ifx [1]{%
 \ifx #1\expandafter \@firstoftwo
 \else \expandafter \@secondoftwo
 \fi
}%
\providecommand \natexlab [1]{#1}%
\providecommand \enquote  [1]{``#1''}%
\providecommand \bibnamefont  [1]{#1}%
\providecommand \bibfnamefont [1]{#1}%
\providecommand \citenamefont [1]{#1}%
\providecommand \href@noop [0]{\@secondoftwo}%
\providecommand \href [0]{\begingroup \@sanitize@url \@href}%
\providecommand \@href[1]{\@@startlink{#1}\@@href}%
\providecommand \@@href[1]{\endgroup#1\@@endlink}%
\providecommand \@sanitize@url [0]{\catcode `\\12\catcode `\$12\catcode
  `\&12\catcode `\#12\catcode `\^12\catcode `\_12\catcode `\%12\relax}%
\providecommand \@@startlink[1]{}%
\providecommand \@@endlink[0]{}%
\providecommand \url  [0]{\begingroup\@sanitize@url \@url }%
\providecommand \@url [1]{\endgroup\@href {#1}{\urlprefix }}%
\providecommand \urlprefix  [0]{URL }%
\providecommand \Eprint [0]{\href }%
\providecommand \doibase [0]{https://doi.org/}%
\providecommand \selectlanguage [0]{\@gobble}%
\providecommand \bibinfo  [0]{\@secondoftwo}%
\providecommand \bibfield  [0]{\@secondoftwo}%
\providecommand \translation [1]{[#1]}%
\providecommand \BibitemOpen [0]{}%
\providecommand \bibitemStop [0]{}%
\providecommand \bibitemNoStop [0]{.\EOS\space}%
\providecommand \EOS [0]{\spacefactor3000\relax}%
\providecommand \BibitemShut  [1]{\csname bibitem#1\endcsname}%
\let\auto@bib@innerbib\@empty
\bibitem [{\citenamefont {Lieb}\ and\ \citenamefont
  {Robinson}(1972)}]{Lieb1972}%
  \BibitemOpen
  \bibfield  {author} {\bibinfo {author} {\bibfnamefont {E.~H.}\ \bibnamefont
  {Lieb}}\ and\ \bibinfo {author} {\bibfnamefont {D.~W.}\ \bibnamefont
  {Robinson}},\ }\bibfield  {title} {\bibinfo {title} {The finite group
  velocity of quantum spin systems},\ }\href
  {https://doi.org/10.1007/BF01645779} {\bibfield  {journal} {\bibinfo
  {journal} {Commun. Math. Phys.}\ }\textbf {\bibinfo {volume} {28}},\ \bibinfo
  {pages} {251} (\bibinfo {year} {1972})}\BibitemShut {NoStop}%
\bibitem [{\citenamefont {Nachtergaele}\ and\ \citenamefont
  {Sims}(2006)}]{NachtergaeleSims2006}%
  \BibitemOpen
  \bibfield  {author} {\bibinfo {author} {\bibfnamefont {B.}~\bibnamefont
  {Nachtergaele}}\ and\ \bibinfo {author} {\bibfnamefont {R.}~\bibnamefont
  {Sims}},\ }\bibfield  {title} {\bibinfo {title} {Lieb-robinson bounds and the
  exponential clustering theorem},\ }\href
  {https://doi.org/10.1007/s00220-006-1556-1} {\bibfield  {journal} {\bibinfo
  {journal} {Communications in Mathematical Physics}\ }\textbf {\bibinfo
  {volume} {265}},\ \bibinfo {pages} {119} (\bibinfo {year} {2006})},\ \Eprint
  {https://arxiv.org/abs/math-ph/0506030} {arXiv:math-ph/0506030} \BibitemShut
  {NoStop}%
\bibitem [{\citenamefont {Hastings}\ and\ \citenamefont
  {Koma}(2006)}]{Hastings2006long}%
  \BibitemOpen
  \bibfield  {author} {\bibinfo {author} {\bibfnamefont {M.~B.}\ \bibnamefont
  {Hastings}}\ and\ \bibinfo {author} {\bibfnamefont {T.}~\bibnamefont
  {Koma}},\ }\bibfield  {title} {\bibinfo {title} {Spectral gap and exponential
  decay of correlations},\ }\href {https://doi.org/10.1007/s00220-006-0030-4}
  {\bibfield  {journal} {\bibinfo  {journal} {Commun. Math. Phys.}\ }\textbf
  {\bibinfo {volume} {265}},\ \bibinfo {pages} {781} (\bibinfo {year}
  {2006})}\BibitemShut {NoStop}%
\bibitem [{\citenamefont {Hastings}(2007)}]{Hastings_2007}%
  \BibitemOpen
  \bibfield  {author} {\bibinfo {author} {\bibfnamefont {M.~B.}\ \bibnamefont
  {Hastings}},\ }\bibfield  {title} {\bibinfo {title} {An area law for
  one-dimensional quantum systems},\ }\href
  {https://doi.org/10.1088/1742-5468/2007/08/P08024} {\bibfield  {journal}
  {\bibinfo  {journal} {J. Stat. Mech.}\ }\textbf {\bibinfo {volume} {2007}},\
  \bibinfo {pages} {P08024} (\bibinfo {year} {2007})}\BibitemShut {NoStop}%
\bibitem [{\citenamefont {Hastings}\ and\ \citenamefont
  {Wen}(2005)}]{Hastings2005adiabatic}%
  \BibitemOpen
  \bibfield  {author} {\bibinfo {author} {\bibfnamefont {M.~B.}\ \bibnamefont
  {Hastings}}\ and\ \bibinfo {author} {\bibfnamefont {X.-G.}\ \bibnamefont
  {Wen}},\ }\bibfield  {title} {\bibinfo {title} {Quasiadiabatic continuation
  of quantum states: The stability of topological ground-state degeneracy and
  emergent gauge invariance},\ }\href
  {https://doi.org/10.1103/PhysRevB.72.045141} {\bibfield  {journal} {\bibinfo
  {journal} {Phys. Rev. B}\ }\textbf {\bibinfo {volume} {72}},\ \bibinfo
  {pages} {045141} (\bibinfo {year} {2005})}\BibitemShut {NoStop}%
\bibitem [{\citenamefont {Bravyi}\ \emph {et~al.}(2006)\citenamefont {Bravyi},
  \citenamefont {Hastings},\ and\ \citenamefont {Verstraete}}]{Bravyi2006}%
  \BibitemOpen
  \bibfield  {author} {\bibinfo {author} {\bibfnamefont {S.}~\bibnamefont
  {Bravyi}}, \bibinfo {author} {\bibfnamefont {M.~B.}\ \bibnamefont
  {Hastings}},\ and\ \bibinfo {author} {\bibfnamefont {F.}~\bibnamefont
  {Verstraete}},\ }\bibfield  {title} {\bibinfo {title} {Lieb-robinson bounds
  and the generation of correlations and topological quantum order},\ }\href
  {https://doi.org/10.1103/PhysRevLett.97.050401} {\bibfield  {journal}
  {\bibinfo  {journal} {Phys. Rev. Lett.}\ }\textbf {\bibinfo {volume} {97}},\
  \bibinfo {pages} {050401} (\bibinfo {year} {2006})}\BibitemShut {NoStop}%
\bibitem [{\citenamefont {Hastings}(2004)}]{Hastings2004LSM}%
  \BibitemOpen
  \bibfield  {author} {\bibinfo {author} {\bibfnamefont {M.~B.}\ \bibnamefont
  {Hastings}},\ }\bibfield  {title} {\bibinfo {title} {Lieb-schultz-mattis in
  higher dimensions},\ }\href {https://doi.org/10.1103/PhysRevB.69.104431}
  {\bibfield  {journal} {\bibinfo  {journal} {Phys. Rev. B}\ }\textbf {\bibinfo
  {volume} {69}},\ \bibinfo {pages} {104431} (\bibinfo {year}
  {2004})}\BibitemShut {NoStop}%
\bibitem [{\citenamefont {Osborne}(2006)}]{Osborne2006}%
  \BibitemOpen
  \bibfield  {author} {\bibinfo {author} {\bibfnamefont {T.~J.}\ \bibnamefont
  {Osborne}},\ }\bibfield  {title} {\bibinfo {title} {Efficient approximation
  of the dynamics of one-dimensional quantum spin systems},\ }\href
  {https://doi.org/10.1103/PhysRevLett.97.157202} {\bibfield  {journal}
  {\bibinfo  {journal} {Phys. Rev. Lett.}\ }\textbf {\bibinfo {volume} {97}},\
  \bibinfo {pages} {157202} (\bibinfo {year} {2006})}\BibitemShut {NoStop}%
\bibitem [{\citenamefont {Haah}\ \emph {et~al.}(2021)\citenamefont {Haah},
  \citenamefont {Hastings}, \citenamefont {Kothari},\ and\ \citenamefont
  {Low}}]{haah2021quantum}%
  \BibitemOpen
  \bibfield  {author} {\bibinfo {author} {\bibfnamefont {J.}~\bibnamefont
  {Haah}}, \bibinfo {author} {\bibfnamefont {M.~B.}\ \bibnamefont {Hastings}},
  \bibinfo {author} {\bibfnamefont {R.}~\bibnamefont {Kothari}},\ and\ \bibinfo
  {author} {\bibfnamefont {G.~H.}\ \bibnamefont {Low}},\ }\bibfield  {title}
  {\bibinfo {title} {{Quantum algorithm for simulating real time evolution of
  lattice Hamiltonians}},\ }\href@noop {} {\bibfield  {journal} {\bibinfo
  {journal} {SIAM Journal on Computing}\ }\textbf {\bibinfo {volume} {52}},\
  \bibinfo {pages} {FOCS18} (\bibinfo {year} {2021})}\BibitemShut {NoStop}%
\bibitem [{\citenamefont {Yu}\ \emph {et~al.}(2025)\citenamefont {Yu},
  \citenamefont {Li}, \citenamefont {Cirac},\ and\ \citenamefont
  {Trivedi}}]{yu2025optimization}%
  \BibitemOpen
  \bibfield  {author} {\bibinfo {author} {\bibfnamefont {X.-H.}\ \bibnamefont
  {Yu}}, \bibinfo {author} {\bibfnamefont {H.}~\bibnamefont {Li}}, \bibinfo
  {author} {\bibfnamefont {J.~I.}\ \bibnamefont {Cirac}},\ and\ \bibinfo
  {author} {\bibfnamefont {R.}~\bibnamefont {Trivedi}},\ }\href
  {https://arxiv.org/abs/2509.02268} {\bibinfo {title} {Optimizing digital
  quantum simulation of open quantum lattice models}} (\bibinfo {year}
  {2025}),\ \Eprint {https://arxiv.org/abs/2509.02268} {arXiv:2509.02268
  [quant-ph]} \BibitemShut {NoStop}%
\bibitem [{\citenamefont {Tran}\ \emph {et~al.}(2019)\citenamefont {Tran},
  \citenamefont {Guo}, \citenamefont {Su}, \citenamefont {Garrison},
  \citenamefont {Eldredge}, \citenamefont {Foss-Feig}, \citenamefont {Childs},\
  and\ \citenamefont {Gorshkov}}]{Tran2019}%
  \BibitemOpen
  \bibfield  {author} {\bibinfo {author} {\bibfnamefont {M.~C.}\ \bibnamefont
  {Tran}}, \bibinfo {author} {\bibfnamefont {A.~Y.}\ \bibnamefont {Guo}},
  \bibinfo {author} {\bibfnamefont {Y.}~\bibnamefont {Su}}, \bibinfo {author}
  {\bibfnamefont {J.~R.}\ \bibnamefont {Garrison}}, \bibinfo {author}
  {\bibfnamefont {Z.}~\bibnamefont {Eldredge}}, \bibinfo {author}
  {\bibfnamefont {M.}~\bibnamefont {Foss-Feig}}, \bibinfo {author}
  {\bibfnamefont {A.~M.}\ \bibnamefont {Childs}},\ and\ \bibinfo {author}
  {\bibfnamefont {A.~V.}\ \bibnamefont {Gorshkov}},\ }\bibfield  {title}
  {\bibinfo {title} {Locality and digital quantum simulation of power-law
  interactions},\ }\href {https://doi.org/10.1103/PhysRevX.9.031006} {\bibfield
   {journal} {\bibinfo  {journal} {Phys. Rev. X}\ }\textbf {\bibinfo {volume}
  {9}},\ \bibinfo {pages} {031006} (\bibinfo {year} {2019})}\BibitemShut
  {NoStop}%
\bibitem [{\citenamefont {Tran}\ \emph {et~al.}(2020)\citenamefont {Tran},
  \citenamefont {Chen}, \citenamefont {Ehrenberg}, \citenamefont {Guo},
  \citenamefont {Deshpande}, \citenamefont {Hong}, \citenamefont {Gong},
  \citenamefont {Gorshkov},\ and\ \citenamefont {Lucas}}]{Tran_hierarchy_2020}%
  \BibitemOpen
  \bibfield  {author} {\bibinfo {author} {\bibfnamefont {M.~C.}\ \bibnamefont
  {Tran}}, \bibinfo {author} {\bibfnamefont {C.-F.}\ \bibnamefont {Chen}},
  \bibinfo {author} {\bibfnamefont {A.}~\bibnamefont {Ehrenberg}}, \bibinfo
  {author} {\bibfnamefont {A.~Y.}\ \bibnamefont {Guo}}, \bibinfo {author}
  {\bibfnamefont {A.}~\bibnamefont {Deshpande}}, \bibinfo {author}
  {\bibfnamefont {Y.}~\bibnamefont {Hong}}, \bibinfo {author} {\bibfnamefont
  {Z.-X.}\ \bibnamefont {Gong}}, \bibinfo {author} {\bibfnamefont {A.~V.}\
  \bibnamefont {Gorshkov}},\ and\ \bibinfo {author} {\bibfnamefont
  {A.}~\bibnamefont {Lucas}},\ }\bibfield  {title} {\bibinfo {title} {Hierarchy
  of linear light cones with long-range interactions},\ }\href
  {https://doi.org/10.1103/PhysRevX.10.031009} {\bibfield  {journal} {\bibinfo
  {journal} {Phys. Rev. X}\ }\textbf {\bibinfo {volume} {10}},\ \bibinfo
  {pages} {031009} (\bibinfo {year} {2020})}\BibitemShut {NoStop}%
\bibitem [{\citenamefont {Kuwahara}\ and\ \citenamefont
  {Saito}(2020)}]{Kuwahara_light2020}%
  \BibitemOpen
  \bibfield  {author} {\bibinfo {author} {\bibfnamefont {T.}~\bibnamefont
  {Kuwahara}}\ and\ \bibinfo {author} {\bibfnamefont {K.}~\bibnamefont
  {Saito}},\ }\bibfield  {title} {\bibinfo {title} {Strictly linear light cones
  in long-range interacting systems of arbitrary dimensions},\ }\href
  {https://doi.org/10.1103/PhysRevX.10.031010} {\bibfield  {journal} {\bibinfo
  {journal} {Phys. Rev. X}\ }\textbf {\bibinfo {volume} {10}},\ \bibinfo
  {pages} {031010} (\bibinfo {year} {2020})}\BibitemShut {NoStop}%
\bibitem [{\citenamefont {Kuwahara}\ and\ \citenamefont
  {Saito}(2021{\natexlab{a}})}]{KuwaharaPRL2021}%
  \BibitemOpen
  \bibfield  {author} {\bibinfo {author} {\bibfnamefont {T.}~\bibnamefont
  {Kuwahara}}\ and\ \bibinfo {author} {\bibfnamefont {K.}~\bibnamefont
  {Saito}},\ }\bibfield  {title} {\bibinfo {title} {Absence of fast scrambling
  in thermodynamically stable long-range interacting systems},\ }\href
  {https://doi.org/10.1103/PhysRevLett.126.030604} {\bibfield  {journal}
  {\bibinfo  {journal} {Phys. Rev. Lett.}\ }\textbf {\bibinfo {volume} {126}},\
  \bibinfo {pages} {030604} (\bibinfo {year} {2021}{\natexlab{a}})}\BibitemShut
  {NoStop}%
\bibitem [{\citenamefont {Chen}\ and\ \citenamefont
  {Lucas}(2019)}]{ChenLucas:2019}%
  \BibitemOpen
  \bibfield  {author} {\bibinfo {author} {\bibfnamefont {C.-F.}\ \bibnamefont
  {Chen}}\ and\ \bibinfo {author} {\bibfnamefont {A.}~\bibnamefont {Lucas}},\
  }\bibfield  {title} {\bibinfo {title} {Finite speed of quantum scrambling
  with long range interactions},\ }\href
  {https://doi.org/10.1103/PhysRevLett.123.250605} {\bibfield  {journal}
  {\bibinfo  {journal} {Phys. Rev. Lett.}\ }\textbf {\bibinfo {volume} {123}},\
  \bibinfo {pages} {250605} (\bibinfo {year} {2019})}\BibitemShut {NoStop}%
\bibitem [{\citenamefont {Colmenarez}\ and\ \citenamefont
  {Luitz}(2020)}]{Colmenarez2020}%
  \BibitemOpen
  \bibfield  {author} {\bibinfo {author} {\bibfnamefont {L.}~\bibnamefont
  {Colmenarez}}\ and\ \bibinfo {author} {\bibfnamefont {D.~J.}\ \bibnamefont
  {Luitz}},\ }\bibfield  {title} {\bibinfo {title} {Lieb-robinson bounds and
  out-of-time order correlators in a long-range spin chain},\ }\href
  {https://doi.org/10.1103/PhysRevResearch.2.043047} {\bibfield  {journal}
  {\bibinfo  {journal} {Phys. Rev. Res.}\ }\textbf {\bibinfo {volume} {2}},\
  \bibinfo {pages} {043047} (\bibinfo {year} {2020})}\BibitemShut {NoStop}%
\bibitem [{\citenamefont {Teufel}(2026)}]{teufel2026lossless}%
  \BibitemOpen
  \bibfield  {author} {\bibinfo {author} {\bibfnamefont {S.}~\bibnamefont
  {Teufel}},\ }\bibfield  {title} {\bibinfo {title} {Lossless propagation
  bounds for long-range fermionic interactions},\ }\href@noop {} {\bibfield
  {journal} {\bibinfo  {journal} {arXiv preprint arXiv:2611.03441}\ } (\bibinfo
  {year} {2026})}\BibitemShut {NoStop}%
\bibitem [{\citenamefont {Kuwahara}\ and\ \citenamefont
  {Saito}(2021{\natexlab{b}})}]{Kuwahara2021}%
  \BibitemOpen
  \bibfield  {author} {\bibinfo {author} {\bibfnamefont {T.}~\bibnamefont
  {Kuwahara}}\ and\ \bibinfo {author} {\bibfnamefont {K.}~\bibnamefont
  {Saito}},\ }\bibfield  {title} {\bibinfo {title} {Lieb-robinson bound and
  almost-linear light cone in interacting boson systems},\ }\href
  {https://doi.org/10.1103/PhysRevLett.127.070403} {\bibfield  {journal}
  {\bibinfo  {journal} {Phys. Rev. Lett.}\ }\textbf {\bibinfo {volume} {127}},\
  \bibinfo {pages} {070403} (\bibinfo {year} {2021}{\natexlab{b}})}\BibitemShut
  {NoStop}%
\bibitem [{\citenamefont {Yin}\ and\ \citenamefont {Lucas}(2022)}]{Yin2022}%
  \BibitemOpen
  \bibfield  {author} {\bibinfo {author} {\bibfnamefont {C.}~\bibnamefont
  {Yin}}\ and\ \bibinfo {author} {\bibfnamefont {A.}~\bibnamefont {Lucas}},\
  }\bibfield  {title} {\bibinfo {title} {Finite speed of quantum information in
  models of interacting bosons at finite density},\ }\href
  {https://doi.org/10.1103/PhysRevX.12.021039} {\bibfield  {journal} {\bibinfo
  {journal} {Phys. Rev. X}\ }\textbf {\bibinfo {volume} {12}},\ \bibinfo
  {pages} {021039} (\bibinfo {year} {2022})}\BibitemShut {NoStop}%
\bibitem [{\citenamefont {Faupin}\ \emph
  {et~al.}(2022{\natexlab{a}})\citenamefont {Faupin}, \citenamefont {Lemm},\
  and\ \citenamefont {Sigal}}]{Faupin_LR2022}%
  \BibitemOpen
  \bibfield  {author} {\bibinfo {author} {\bibfnamefont {J.}~\bibnamefont
  {Faupin}}, \bibinfo {author} {\bibfnamefont {M.}~\bibnamefont {Lemm}},\ and\
  \bibinfo {author} {\bibfnamefont {I.~M.}\ \bibnamefont {Sigal}},\ }\bibfield
  {title} {\bibinfo {title} {On lieb--robinson bounds for the bose--hubbard
  model},\ }\href {https://doi.org/10.1007/s00220-022-04416-8} {\bibfield
  {journal} {\bibinfo  {journal} {Commun. Math. Phys.}\ }\textbf {\bibinfo
  {volume} {394}},\ \bibinfo {pages} {1011} (\bibinfo {year}
  {2022}{\natexlab{a}})}\BibitemShut {NoStop}%
\bibitem [{\citenamefont {Lemm}\ \emph
  {et~al.}(2023{\natexlab{a}})\citenamefont {Lemm}, \citenamefont {Rubiliani},
  \citenamefont {Sigal},\ and\ \citenamefont {Zhang}}]{Lemm2023info}%
  \BibitemOpen
  \bibfield  {author} {\bibinfo {author} {\bibfnamefont {M.}~\bibnamefont
  {Lemm}}, \bibinfo {author} {\bibfnamefont {C.}~\bibnamefont {Rubiliani}},
  \bibinfo {author} {\bibfnamefont {I.~M.}\ \bibnamefont {Sigal}},\ and\
  \bibinfo {author} {\bibfnamefont {J.}~\bibnamefont {Zhang}},\ }\bibfield
  {title} {\bibinfo {title} {Information propagation in long-range quantum
  many-body systems},\ }\href {https://doi.org/10.1103/PhysRevA.108.L060401}
  {\bibfield  {journal} {\bibinfo  {journal} {Phys. Rev. A}\ }\textbf {\bibinfo
  {volume} {108}},\ \bibinfo {pages} {L060401} (\bibinfo {year}
  {2023}{\natexlab{a}})}\BibitemShut {NoStop}%
\bibitem [{\citenamefont {Kuwahara}\ \emph {et~al.}(2024)\citenamefont
  {Kuwahara}, \citenamefont {Van~Vu},\ and\ \citenamefont
  {Saito}}]{Kuwahara2024}%
  \BibitemOpen
  \bibfield  {author} {\bibinfo {author} {\bibfnamefont {T.}~\bibnamefont
  {Kuwahara}}, \bibinfo {author} {\bibfnamefont {T.}~\bibnamefont {Van~Vu}},\
  and\ \bibinfo {author} {\bibfnamefont {K.}~\bibnamefont {Saito}},\ }\bibfield
   {title} {\bibinfo {title} {Effective light cone and digital quantum
  simulation of interacting bosons},\ }\href
  {https://doi.org/10.1038/s41467-024-46501-7} {\bibfield  {journal} {\bibinfo
  {journal} {Nat. Commun.}\ }\textbf {\bibinfo {volume} {15}},\ \bibinfo
  {pages} {2520} (\bibinfo {year} {2024})}\BibitemShut {NoStop}%
\bibitem [{\citenamefont {Elgart}\ and\ \citenamefont
  {Lemm}(2026)}]{elgart2026lieb}%
  \BibitemOpen
  \bibfield  {author} {\bibinfo {author} {\bibfnamefont {A.}~\bibnamefont
  {Elgart}}\ and\ \bibinfo {author} {\bibfnamefont {M.}~\bibnamefont {Lemm}},\
  }\bibfield  {title} {\bibinfo {title} {{Lieb-Robinson bounds for Bose-Hubbard
  Hamiltonians via path expansions}},\ }\href
  {https://arxiv.org/abs/2611.02263} {\bibfield  {journal} {\bibinfo  {journal}
  {arXiv preprint arXiv:2611.02263}\ } (\bibinfo {year} {2026})}\BibitemShut
  {NoStop}%
\bibitem [{\citenamefont {Kuwahara}\ and\ \citenamefont
  {Lemm}(2026)}]{KuwaharaLemm2026}%
  \BibitemOpen
  \bibfield  {author} {\bibinfo {author} {\bibfnamefont {T.}~\bibnamefont
  {Kuwahara}}\ and\ \bibinfo {author} {\bibfnamefont {M.}~\bibnamefont
  {Lemm}},\ }\bibfield  {title} {\bibinfo {title} {Symmetry-enhanced
  {Lieb--Robinson} bounds for a class of {Bose--Hubbard} type hamiltonians},\
  }\bibfield  {journal} {\bibinfo  {journal} {Annales Henri Poincar{\'e}}\
  }\href {https://doi.org/10.1007/s00023-026-01733-9}
  {10.1007/s00023-026-01733-9} (\bibinfo {year} {2026}),\ \Eprint
  {https://arxiv.org/abs/2405.04672} {arXiv:2405.04672 [math-ph]} \BibitemShut
  {NoStop}%
\bibitem [{\citenamefont {Lemm}\ and\ \citenamefont
  {Rubiliani}(2026)}]{Carlalemm2026liebrobinsonbounds}%
  \BibitemOpen
  \bibfield  {author} {\bibinfo {author} {\bibfnamefont {M.}~\bibnamefont
  {Lemm}}\ and\ \bibinfo {author} {\bibfnamefont {C.}~\bibnamefont
  {Rubiliani}},\ }\href {https://arxiv.org/abs/2603.26209} {\bibinfo {title}
  {Lieb-robinson bounds for bose-hubbard hamiltonians: A review with a
  simplified proof}} (\bibinfo {year} {2026}),\ \Eprint
  {https://arxiv.org/abs/2603.26209} {arXiv:2603.26209 [math-ph]} \BibitemShut
  {NoStop}%
\bibitem [{\citenamefont {Rubiliani}\ \emph
  {et~al.}(2026{\natexlab{a}})\citenamefont {Rubiliani}, \citenamefont
  {Shang},\ and\ \citenamefont {Kuwahara}}]{rubiliani2026lieb}%
  \BibitemOpen
  \bibfield  {author} {\bibinfo {author} {\bibfnamefont {C.}~\bibnamefont
  {Rubiliani}}, \bibinfo {author} {\bibfnamefont {C.}~\bibnamefont {Shang}},\
  and\ \bibinfo {author} {\bibfnamefont {T.}~\bibnamefont {Kuwahara}},\
  }\bibfield  {title} {\bibinfo {title} {{Lieb-Robinson Light Cones in
  Long-Range Bosonic Systems}},\ }\href {https://arxiv.org/abs/2611.00776}
  {\bibfield  {journal} {\bibinfo  {journal} {arXiv preprint arXiv:2611.00776}\
  } (\bibinfo {year} {2026}{\natexlab{a}})}\BibitemShut {NoStop}%
\bibitem [{\citenamefont {Poulin}(2010)}]{LRopen_2010}%
  \BibitemOpen
  \bibfield  {author} {\bibinfo {author} {\bibfnamefont {D.}~\bibnamefont
  {Poulin}},\ }\bibfield  {title} {\bibinfo {title} {Lieb-robinson bound and
  locality for general markovian quantum dynamics},\ }\href
  {https://doi.org/10.1103/PhysRevLett.104.190401} {\bibfield  {journal}
  {\bibinfo  {journal} {Phys. Rev. Lett.}\ }\textbf {\bibinfo {volume} {104}},\
  \bibinfo {pages} {190401} (\bibinfo {year} {2010})}\BibitemShut {NoStop}%
\bibitem [{\citenamefont {Barthel}\ and\ \citenamefont
  {Kliesch}(2012)}]{Barthel2012}%
  \BibitemOpen
  \bibfield  {author} {\bibinfo {author} {\bibfnamefont {T.}~\bibnamefont
  {Barthel}}\ and\ \bibinfo {author} {\bibfnamefont {M.}~\bibnamefont
  {Kliesch}},\ }\bibfield  {title} {\bibinfo {title} {Quasilocality and
  efficient simulation of markovian quantum dynamics},\ }\href
  {https://doi.org/10.1103/PhysRevLett.108.230504} {\bibfield  {journal}
  {\bibinfo  {journal} {Phys. Rev. Lett.}\ }\textbf {\bibinfo {volume} {108}},\
  \bibinfo {pages} {230504} (\bibinfo {year} {2012})}\BibitemShut {NoStop}%
\bibitem [{\citenamefont {Han}\ and\ \citenamefont
  {Hartnoll}(2018)}]{XHan2018}%
  \BibitemOpen
  \bibfield  {author} {\bibinfo {author} {\bibfnamefont {X.}~\bibnamefont
  {Han}}\ and\ \bibinfo {author} {\bibfnamefont {S.~A.}\ \bibnamefont
  {Hartnoll}},\ }\bibfield  {title} {\bibinfo {title} {Locality bound for
  dissipative quantum transport},\ }\href
  {https://doi.org/10.1103/PhysRevLett.121.170601} {\bibfield  {journal}
  {\bibinfo  {journal} {Phys. Rev. Lett.}\ }\textbf {\bibinfo {volume} {121}},\
  \bibinfo {pages} {170601} (\bibinfo {year} {2018})}\BibitemShut {NoStop}%
\bibitem [{\citenamefont {Sweke}\ \emph {et~al.}(2019)\citenamefont {Sweke},
  \citenamefont {Eisert},\ and\ \citenamefont {Kastner}}]{Sweke_2019}%
  \BibitemOpen
  \bibfield  {author} {\bibinfo {author} {\bibfnamefont {R.}~\bibnamefont
  {Sweke}}, \bibinfo {author} {\bibfnamefont {J.}~\bibnamefont {Eisert}},\ and\
  \bibinfo {author} {\bibfnamefont {M.}~\bibnamefont {Kastner}},\ }\bibfield
  {title} {\bibinfo {title} {Lieb–robinson bounds for open quantum systems
  with long-ranged interactions},\ }\href
  {https://doi.org/10.1088/1751-8121/ab3f4a} {\bibfield  {journal} {\bibinfo
  {journal} {J. Phys. A}\ }\textbf {\bibinfo {volume} {52}},\ \bibinfo {pages}
  {424003} (\bibinfo {year} {2019})}\BibitemShut {NoStop}%
\bibitem [{\citenamefont {Li}\ \emph {et~al.}(2025)\citenamefont {Li},
  \citenamefont {Yu}, \citenamefont {Nakagawa},\ and\ \citenamefont
  {Ueda}}]{Li2024BEC}%
  \BibitemOpen
  \bibfield  {author} {\bibinfo {author} {\bibfnamefont {H.}~\bibnamefont
  {Li}}, \bibinfo {author} {\bibfnamefont {X.-H.}\ \bibnamefont {Yu}}, \bibinfo
  {author} {\bibfnamefont {M.}~\bibnamefont {Nakagawa}},\ and\ \bibinfo
  {author} {\bibfnamefont {M.}~\bibnamefont {Ueda}},\ }\bibfield  {title}
  {\bibinfo {title} {Dissipative superfluidity in a molecular bose-einstein
  condensate},\ }\href {https://doi.org/10.1103/ww8r-sjhb} {\bibfield
  {journal} {\bibinfo  {journal} {Phys. Rev. Lett.}\ }\textbf {\bibinfo
  {volume} {135}},\ \bibinfo {pages} {166001} (\bibinfo {year}
  {2025})}\BibitemShut {NoStop}%
\bibitem [{\citenamefont {Trivedi}\ \emph {et~al.}(2025)\citenamefont
  {Trivedi}, \citenamefont {Yu},\ and\ \citenamefont {Rudner}}]{Rahul2024}%
  \BibitemOpen
  \bibfield  {author} {\bibinfo {author} {\bibfnamefont {R.}~\bibnamefont
  {Trivedi}}, \bibinfo {author} {\bibfnamefont {X.}~\bibnamefont {Yu}},\ and\
  \bibinfo {author} {\bibfnamefont {M.}~\bibnamefont {Rudner}},\ }\href
  {https://arxiv.org/abs/2410.15481} {\bibinfo {title} {A lieb-robinson bound
  for open quantum systems with memory}} (\bibinfo {year} {2025}),\ \Eprint
  {https://arxiv.org/abs/2410.15481} {arXiv:2410.15481 [quant-ph]} \BibitemShut
  {NoStop}%
\bibitem [{\citenamefont {Lemm}\ and\ \citenamefont
  {M{\"o}bus}(2026)}]{lemm2026dissipation}%
  \BibitemOpen
  \bibfield  {author} {\bibinfo {author} {\bibfnamefont {M.}~\bibnamefont
  {Lemm}}\ and\ \bibinfo {author} {\bibfnamefont {T.}~\bibnamefont
  {M{\"o}bus}},\ }\bibfield  {title} {\bibinfo {title} {{Dissipation-enabled
  operator-norm locality bounds for bose-hubbard hamiltonians}},\ }\href
  {https://arxiv.org/abs/2610.01669} {\bibfield  {journal} {\bibinfo  {journal}
  {arXiv preprint arXiv:2610.01669}\ } (\bibinfo {year} {2026})}\BibitemShut
  {NoStop}%
\bibitem [{\citenamefont {Rubiliani}\ \emph
  {et~al.}(2026{\natexlab{b}})\citenamefont {Rubiliani}, \citenamefont
  {Wesle},\ and\ \citenamefont {Wessel}}]{rubiliani2026lieb_open}%
  \BibitemOpen
  \bibfield  {author} {\bibinfo {author} {\bibfnamefont {C.}~\bibnamefont
  {Rubiliani}}, \bibinfo {author} {\bibfnamefont {M.}~\bibnamefont {Wesle}},\
  and\ \bibinfo {author} {\bibfnamefont {T.}~\bibnamefont {Wessel}},\
  }\bibfield  {title} {\bibinfo {title} {Lieb-robinson bounds for irreversible
  dynamics via adaptive dyson expansion},\ }\href
  {https://arxiv.org/abs/2610.12731} {\bibfield  {journal} {\bibinfo  {journal}
  {arXiv preprint arXiv:2610.12731}\ } (\bibinfo {year}
  {2026}{\natexlab{b}})}\BibitemShut {NoStop}%
\bibitem [{\citenamefont {Lemm}\ \emph {et~al.}(2026)\citenamefont {Lemm},
  \citenamefont {Lu},\ and\ \citenamefont {Singh}}]{lemm2026lieb_continuum}%
  \BibitemOpen
  \bibfield  {author} {\bibinfo {author} {\bibfnamefont {M.}~\bibnamefont
  {Lemm}}, \bibinfo {author} {\bibfnamefont {H.}~\bibnamefont {Lu}},\ and\
  \bibinfo {author} {\bibfnamefont {H.~P.}\ \bibnamefont {Singh}},\ }\bibfield
  {title} {\bibinfo {title} {Lieb-robinson bound for one-dimensional continuum
  fermions},\ }\href {https://arxiv.org/abs/2611.11655} {\bibfield  {journal}
  {\bibinfo  {journal} {arXiv preprint arXiv:2611.11655}\ } (\bibinfo {year}
  {2026})}\BibitemShut {NoStop}%
\bibitem [{\citenamefont {Hinrichs}\ \emph {et~al.}(2025)\citenamefont
  {Hinrichs}, \citenamefont {Lemm},\ and\ \citenamefont
  {Siebert}}]{Hinrichs2025}%
  \BibitemOpen
  \bibfield  {author} {\bibinfo {author} {\bibfnamefont {B.}~\bibnamefont
  {Hinrichs}}, \bibinfo {author} {\bibfnamefont {M.}~\bibnamefont {Lemm}},\
  and\ \bibinfo {author} {\bibfnamefont {O.}~\bibnamefont {Siebert}},\
  }\bibfield  {title} {\bibinfo {title} {On lieb-robinson bounds for a class of
  continuum fermions},\ }\href {https://doi.org/10.1007/s00023-024-01453-y}
  {\bibfield  {journal} {\bibinfo  {journal} {Annales Henri Poincar{\'e}}\
  }\textbf {\bibinfo {volume} {26}},\ \bibinfo {pages} {41} (\bibinfo {year}
  {2025})}\BibitemShut {NoStop}%
\bibitem [{\citenamefont {Bachmann}\ and\ \citenamefont
  {De~Nittis}(2025)}]{Bachmann2025}%
  \BibitemOpen
  \bibfield  {author} {\bibinfo {author} {\bibfnamefont {S.}~\bibnamefont
  {Bachmann}}\ and\ \bibinfo {author} {\bibfnamefont {G.}~\bibnamefont
  {De~Nittis}},\ }\bibfield  {title} {\bibinfo {title} {Lieb-robinson bounds in
  the continuum via localized frames},\ }\href
  {https://doi.org/10.1007/s00023-024-01511-5} {\bibfield  {journal} {\bibinfo
  {journal} {Annales Henri Poincar{\'e}}\ }\textbf {\bibinfo {volume} {26}},\
  \bibinfo {pages} {1} (\bibinfo {year} {2025})}\BibitemShut {NoStop}%
\bibitem [{\citenamefont {Gebert}\ \emph {et~al.}(2020)\citenamefont {Gebert},
  \citenamefont {Nachtergaele}, \citenamefont {Reschke},\ and\ \citenamefont
  {Sims}}]{gebert2020lieb}%
  \BibitemOpen
  \bibfield  {author} {\bibinfo {author} {\bibfnamefont {M.}~\bibnamefont
  {Gebert}}, \bibinfo {author} {\bibfnamefont {B.}~\bibnamefont
  {Nachtergaele}}, \bibinfo {author} {\bibfnamefont {J.}~\bibnamefont
  {Reschke}},\ and\ \bibinfo {author} {\bibfnamefont {R.}~\bibnamefont
  {Sims}},\ }\bibfield  {title} {\bibinfo {title} {Lieb--robinson bounds and
  strongly continuous dynamics for a class of many-body fermion systems in
  $\mathbb{R}^d$},\ }\href {https://doi.org/10.1007/s00023-020-00959-5}
  {\bibfield  {journal} {\bibinfo  {journal} {Annales Henri Poincar{\'e}}\
  }\textbf {\bibinfo {volume} {21}},\ \bibinfo {pages} {3609} (\bibinfo {year}
  {2020})}\BibitemShut {NoStop}%
\bibitem [{\citenamefont {Faupin}\ \emph
  {et~al.}(2022{\natexlab{b}})\citenamefont {Faupin}, \citenamefont {Lemm},\
  and\ \citenamefont {Sigal}}]{Faupin2022}%
  \BibitemOpen
  \bibfield  {author} {\bibinfo {author} {\bibfnamefont {J.}~\bibnamefont
  {Faupin}}, \bibinfo {author} {\bibfnamefont {M.}~\bibnamefont {Lemm}},\ and\
  \bibinfo {author} {\bibfnamefont {I.~M.}\ \bibnamefont {Sigal}},\ }\bibfield
  {title} {\bibinfo {title} {Maximal speed for macroscopic particle transport
  in the bose-hubbard model},\ }\href
  {https://doi.org/10.1103/PhysRevLett.128.150602} {\bibfield  {journal}
  {\bibinfo  {journal} {Phys. Rev. Lett.}\ }\textbf {\bibinfo {volume} {128}},\
  \bibinfo {pages} {150602} (\bibinfo {year} {2022}{\natexlab{b}})}\BibitemShut
  {NoStop}%
\bibitem [{\citenamefont {Van~Vu}\ \emph {et~al.}(2024)\citenamefont {Van~Vu},
  \citenamefont {Kuwahara},\ and\ \citenamefont {Saito}}]{Vu_2024}%
  \BibitemOpen
  \bibfield  {author} {\bibinfo {author} {\bibfnamefont {T.}~\bibnamefont
  {Van~Vu}}, \bibinfo {author} {\bibfnamefont {T.}~\bibnamefont {Kuwahara}},\
  and\ \bibinfo {author} {\bibfnamefont {K.}~\bibnamefont {Saito}},\ }\bibfield
   {title} {\bibinfo {title} {Optimal light cone for macroscopic particle
  transport in long-range systems: A quantum speed limit approach},\ }\href
  {https://doi.org/10.22331/q-2024-09-25-1483} {\bibfield  {journal} {\bibinfo
  {journal} {Quantum}\ }\textbf {\bibinfo {volume} {8}},\ \bibinfo {pages}
  {1483} (\bibinfo {year} {2024})}\BibitemShut {NoStop}%
\bibitem [{\citenamefont {Li}\ \emph {et~al.}(2026)\citenamefont {Li},
  \citenamefont {Shang}, \citenamefont {Kuwahara},\ and\ \citenamefont
  {Vu}}]{Hongchao2026}%
  \BibitemOpen
  \bibfield  {author} {\bibinfo {author} {\bibfnamefont {H.}~\bibnamefont
  {Li}}, \bibinfo {author} {\bibfnamefont {C.}~\bibnamefont {Shang}}, \bibinfo
  {author} {\bibfnamefont {T.}~\bibnamefont {Kuwahara}},\ and\ \bibinfo
  {author} {\bibfnamefont {T.~V.}\ \bibnamefont {Vu}},\ }\bibfield  {title}
  {\bibinfo {title} {Macroscopic particle transport in dissipative long-range
  bosonic systems},\ }\href {https://doi.org/10.1038/s41467-026-70881-7}
  {\bibfield  {journal} {\bibinfo  {journal} {Nature Communications}\ }\textbf
  {\bibinfo {volume} {17}},\ \bibinfo {pages} {4289} (\bibinfo {year}
  {2026})}\BibitemShut {NoStop}%
\bibitem [{\citenamefont {Lemm}\ \emph
  {et~al.}(2023{\natexlab{b}})\citenamefont {Lemm}, \citenamefont {Rubiliani},\
  and\ \citenamefont {Zhang}}]{lemm2023microscopic}%
  \BibitemOpen
  \bibfield  {author} {\bibinfo {author} {\bibfnamefont {M.}~\bibnamefont
  {Lemm}}, \bibinfo {author} {\bibfnamefont {C.}~\bibnamefont {Rubiliani}},\
  and\ \bibinfo {author} {\bibfnamefont {J.}~\bibnamefont {Zhang}},\ }\bibfield
   {title} {\bibinfo {title} {{On the microscopic propagation speed of
  long-range quantum many-body systems}},\ }\href
  {https://arxiv.org/abs/2310.14896} {\bibfield  {journal} {\bibinfo  {journal}
  {arXiv:2310.14896}\ } (\bibinfo {year} {2023}{\natexlab{b}})}\BibitemShut
  {NoStop}%
\bibitem [{\citenamefont {Faupin}\ \emph {et~al.}(2025)\citenamefont {Faupin},
  \citenamefont {Lemm}, \citenamefont {Sigal},\ and\ \citenamefont
  {Zhang}}]{Faupin2025}%
  \BibitemOpen
  \bibfield  {author} {\bibinfo {author} {\bibfnamefont {J.}~\bibnamefont
  {Faupin}}, \bibinfo {author} {\bibfnamefont {M.}~\bibnamefont {Lemm}},
  \bibinfo {author} {\bibfnamefont {I.~M.}\ \bibnamefont {Sigal}},\ and\
  \bibinfo {author} {\bibfnamefont {J.}~\bibnamefont {Zhang}},\ }\bibfield
  {title} {\bibinfo {title} {Macroscopic suppression of supersonic quantum
  transport},\ }\href {https://doi.org/10.1103/27qs-vlrn} {\bibfield  {journal}
  {\bibinfo  {journal} {Phys. Rev. Lett.}\ }\textbf {\bibinfo {volume} {135}},\
  \bibinfo {pages} {160405} (\bibinfo {year} {2025})}\BibitemShut {NoStop}%
\bibitem [{\citenamefont {Kastner}(2011)}]{kastner2011diverging}%
  \BibitemOpen
  \bibfield  {author} {\bibinfo {author} {\bibfnamefont {M.}~\bibnamefont
  {Kastner}},\ }\bibfield  {title} {\bibinfo {title} {Diverging equilibration
  times in long-range quantum spin models},\ }\href
  {https://doi.org/10.1103/PhysRevLett.106.130601} {\bibfield  {journal}
  {\bibinfo  {journal} {Phys. Rev. Lett.}\ }\textbf {\bibinfo {volume} {106}},\
  \bibinfo {pages} {130601} (\bibinfo {year} {2011})}\BibitemShut {NoStop}%
\bibitem [{\citenamefont {Neyenhuis}\ \emph {et~al.}(2017)\citenamefont
  {Neyenhuis}, \citenamefont {Zhang}, \citenamefont {Hess}, \citenamefont
  {Smith}, \citenamefont {Lee}, \citenamefont {Richerme}, \citenamefont {Gong},
  \citenamefont {Gorshkov},\ and\ \citenamefont
  {Monroe}}]{Neyenhuis2017Prethermalization}%
  \BibitemOpen
  \bibfield  {author} {\bibinfo {author} {\bibfnamefont {B.}~\bibnamefont
  {Neyenhuis}}, \bibinfo {author} {\bibfnamefont {J.}~\bibnamefont {Zhang}},
  \bibinfo {author} {\bibfnamefont {P.~W.}\ \bibnamefont {Hess}}, \bibinfo
  {author} {\bibfnamefont {J.}~\bibnamefont {Smith}}, \bibinfo {author}
  {\bibfnamefont {A.~C.}\ \bibnamefont {Lee}}, \bibinfo {author} {\bibfnamefont
  {P.}~\bibnamefont {Richerme}}, \bibinfo {author} {\bibfnamefont {Z.-X.}\
  \bibnamefont {Gong}}, \bibinfo {author} {\bibfnamefont {A.~V.}\ \bibnamefont
  {Gorshkov}},\ and\ \bibinfo {author} {\bibfnamefont {C.}~\bibnamefont
  {Monroe}},\ }\bibfield  {title} {\bibinfo {title} {Observation of
  prethermalization in long-range interacting spin chains},\ }\href
  {https://doi.org/10.1126/sciadv.1700672} {\bibfield  {journal} {\bibinfo
  {journal} {Science Advances}\ }\textbf {\bibinfo {volume} {3}},\ \bibinfo
  {pages} {e1700672} (\bibinfo {year} {2017})}\BibitemShut {NoStop}%
\bibitem [{\citenamefont {Monthus}\ and\ \citenamefont
  {Garel}(2013)}]{MonthusGarel2013DysonBarriers}%
  \BibitemOpen
  \bibfield  {author} {\bibinfo {author} {\bibfnamefont {C.}~\bibnamefont
  {Monthus}}\ and\ \bibinfo {author} {\bibfnamefont {T.}~\bibnamefont
  {Garel}},\ }\bibfield  {title} {\bibinfo {title} {Dynamical barriers in the
  {Dyson} hierarchical model via real space renormalization},\ }\href
  {https://doi.org/10.1088/1742-5468/2013/02/P02023} {\bibfield  {journal}
  {\bibinfo  {journal} {Journal of Statistical Mechanics: Theory and
  Experiment}\ }\textbf {\bibinfo {volume} {2013}},\ \bibinfo {pages} {P02023}
  (\bibinfo {year} {2013})}\BibitemShut {NoStop}%
\bibitem [{\citenamefont {Lerose}\ \emph {et~al.}(2019)\citenamefont {Lerose},
  \citenamefont {{\ifmmode \check{Z}\else \v{Z}\fi{}}unkovi{\ifmmode
  \check{c}\else \v{c}\fi{}}}, \citenamefont {Silva},\ and\ \citenamefont
  {Gambassi}}]{Lenrose2019}%
  \BibitemOpen
  \bibfield  {author} {\bibinfo {author} {\bibfnamefont {A.}~\bibnamefont
  {Lerose}}, \bibinfo {author} {\bibfnamefont {B.}~\bibnamefont {{\ifmmode
  \check{Z}\else \v{Z}\fi{}}unkovi{\ifmmode \check{c}\else \v{c}\fi{}}}},
  \bibinfo {author} {\bibfnamefont {A.}~\bibnamefont {Silva}},\ and\ \bibinfo
  {author} {\bibfnamefont {A.}~\bibnamefont {Gambassi}},\ }\bibfield  {title}
  {\bibinfo {title} {Quasilocalized excitations induced by long-range
  interactions in translationally invariant quantum spin chains},\ }\href
  {https://doi.org/10.1103/PhysRevB.99.121112} {\bibfield  {journal} {\bibinfo
  {journal} {Phys. Rev. B}\ }\textbf {\bibinfo {volume} {99}},\ \bibinfo
  {pages} {121112(R)} (\bibinfo {year} {2019})}\BibitemShut {NoStop}%
\bibitem [{\citenamefont {Liu}\ \emph {et~al.}(2019)\citenamefont {Liu},
  \citenamefont {Lundgren}, \citenamefont {Titum}, \citenamefont {Pagano},
  \citenamefont {Zhang}, \citenamefont {Monroe},\ and\ \citenamefont
  {Gorshkov}}]{liu2019confined}%
  \BibitemOpen
  \bibfield  {author} {\bibinfo {author} {\bibfnamefont {F.}~\bibnamefont
  {Liu}}, \bibinfo {author} {\bibfnamefont {R.}~\bibnamefont {Lundgren}},
  \bibinfo {author} {\bibfnamefont {P.}~\bibnamefont {Titum}}, \bibinfo
  {author} {\bibfnamefont {G.}~\bibnamefont {Pagano}}, \bibinfo {author}
  {\bibfnamefont {J.}~\bibnamefont {Zhang}}, \bibinfo {author} {\bibfnamefont
  {C.}~\bibnamefont {Monroe}},\ and\ \bibinfo {author} {\bibfnamefont {A.~V.}\
  \bibnamefont {Gorshkov}},\ }\bibfield  {title} {\bibinfo {title} {Confined
  quasiparticle dynamics in long-range interacting quantum spin chains},\
  }\href {https://doi.org/10.1103/PhysRevLett.122.150601} {\bibfield  {journal}
  {\bibinfo  {journal} {Phys. Rev. Lett.}\ }\textbf {\bibinfo {volume} {122}},\
  \bibinfo {pages} {150601} (\bibinfo {year} {2019})}\BibitemShut {NoStop}%
\bibitem [{Note1()}]{Note1}%
  \BibitemOpen
  \bibinfo {note} {For time-dependent Hamiltonians, the unitary time evolution
  is governed by the time-ordered exponential $\protect \mathcal {T}\exp
  [-\protect \mathrm {i}\intop \ilimits@ _0^tH(s)\protect \,ds]$. The proof is
  exactly the same because all we use is that the upper bound on the hopping
  coefficients $|J_{ij}(t)|$ is time-independent.}\BibitemShut {Stop}%
\bibitem [{Sup()}]{SupplementaryMaterial}%
  \BibitemOpen
  \href@noop {} {\bibinfo {title} {See supplemental material for
  details.}}\BibitemShut {Stop}%
\bibitem [{\citenamefont {Defenu}\ \emph {et~al.}(2023)\citenamefont {Defenu},
  \citenamefont {Donner}, \citenamefont {Macr\`{\i}}, \citenamefont {Pagano},
  \citenamefont {Ruffo},\ and\ \citenamefont {Trombettoni}}]{Defenu_RMP_2023}%
  \BibitemOpen
  \bibfield  {author} {\bibinfo {author} {\bibfnamefont {N.}~\bibnamefont
  {Defenu}}, \bibinfo {author} {\bibfnamefont {T.}~\bibnamefont {Donner}},
  \bibinfo {author} {\bibfnamefont {T.}~\bibnamefont {Macr\`{\i}}}, \bibinfo
  {author} {\bibfnamefont {G.}~\bibnamefont {Pagano}}, \bibinfo {author}
  {\bibfnamefont {S.}~\bibnamefont {Ruffo}},\ and\ \bibinfo {author}
  {\bibfnamefont {A.}~\bibnamefont {Trombettoni}},\ }\bibfield  {title}
  {\bibinfo {title} {Long-range interacting quantum systems},\ }\href
  {https://doi.org/10.1103/RevModPhys.95.035002} {\bibfield  {journal}
  {\bibinfo  {journal} {Rev. Mod. Phys.}\ }\textbf {\bibinfo {volume} {95}},\
  \bibinfo {pages} {035002} (\bibinfo {year} {2023})}\BibitemShut {NoStop}%
\bibitem [{\citenamefont {Hara}\ and\ \citenamefont
  {Koike}(2026)}]{HaraKoike2026}%
  \BibitemOpen
  \bibfield  {author} {\bibinfo {author} {\bibfnamefont {T.}~\bibnamefont
  {Hara}}\ and\ \bibinfo {author} {\bibfnamefont {T.}~\bibnamefont {Koike}},\
  }\bibfield  {title} {\bibinfo {title} {Timescale for macroscopic
  equilibration in isolated quantum systems: a rigorous derivation for free
  fermions},\ }\href {https://doi.org/10.1103/xxxk-y6yp} {\bibfield  {journal}
  {\bibinfo  {journal} {Physical Review Letters}\ }\textbf {\bibinfo {volume}
  {137}},\ \bibinfo {pages} {030406} (\bibinfo {year} {2026})},\ \bibinfo
  {note} {version 2},\ \Eprint {https://arxiv.org/abs/2602.14096}
  {arXiv:2602.14096 [math-ph]} \BibitemShut {NoStop}%
\bibitem [{\citenamefont {Sigal}\ and\ \citenamefont {Wu}(2025)}]{Sigallr2025}%
  \BibitemOpen
  \bibfield  {author} {\bibinfo {author} {\bibfnamefont {I.~M.}\ \bibnamefont
  {Sigal}}\ and\ \bibinfo {author} {\bibfnamefont {X.}~\bibnamefont {Wu}},\
  }\bibfield  {title} {\bibinfo {title} {On propagation of information in
  quantum mechanics and maximal velocity bounds},\ }\href
  {https://doi.org/10.1007/s11005-025-01899-z} {\bibfield  {journal} {\bibinfo
  {journal} {Letters in Mathematical Physics}\ }\textbf {\bibinfo {volume}
  {115}},\ \bibinfo {pages} {17} (\bibinfo {year} {2025})}\BibitemShut
  {NoStop}%
\bibitem [{\citenamefont {Bloch}(2005)}]{bloch2005ultracold}%
  \BibitemOpen
  \bibfield  {author} {\bibinfo {author} {\bibfnamefont {I.}~\bibnamefont
  {Bloch}},\ }\bibfield  {title} {\bibinfo {title} {{Ultracold quantum gases in
  optical lattices}},\ }\href {https://doi.org//10.1038/nphys138} {\bibfield
  {journal} {\bibinfo  {journal} {Nat. Phys.}\ }\textbf {\bibinfo {volume}
  {1}},\ \bibinfo {pages} {23} (\bibinfo {year} {2005})}\BibitemShut {NoStop}%
\bibitem [{\citenamefont {Macr\`{\i}}\ and\ \citenamefont
  {Pohl}(2014)}]{Macri2014}%
  \BibitemOpen
  \bibfield  {author} {\bibinfo {author} {\bibfnamefont {T.}~\bibnamefont
  {Macr\`{\i}}}\ and\ \bibinfo {author} {\bibfnamefont {T.}~\bibnamefont
  {Pohl}},\ }\bibfield  {title} {\bibinfo {title} {Rydberg dressing of atoms in
  optical lattices},\ }\href {https://doi.org/10.1103/PhysRevA.89.011402}
  {\bibfield  {journal} {\bibinfo  {journal} {Phys. Rev. A}\ }\textbf {\bibinfo
  {volume} {89}},\ \bibinfo {pages} {011402} (\bibinfo {year}
  {2014})}\BibitemShut {NoStop}%
\bibitem [{\citenamefont {Hollerith}\ \emph {et~al.}(2022)\citenamefont
  {Hollerith}, \citenamefont {Srakaew}, \citenamefont {Wei}, \citenamefont
  {Rubio-Abadal}, \citenamefont {Adler}, \citenamefont {Weckesser},
  \citenamefont {Kruckenhauser}, \citenamefont {Walther}, \citenamefont {van
  Bijnen}, \citenamefont {Rui}, \citenamefont {Gross}, \citenamefont {Bloch},\
  and\ \citenamefont {Zeiher}}]{Hollerith2022PRL}%
  \BibitemOpen
  \bibfield  {author} {\bibinfo {author} {\bibfnamefont {S.}~\bibnamefont
  {Hollerith}}, \bibinfo {author} {\bibfnamefont {K.}~\bibnamefont {Srakaew}},
  \bibinfo {author} {\bibfnamefont {D.}~\bibnamefont {Wei}}, \bibinfo {author}
  {\bibfnamefont {A.}~\bibnamefont {Rubio-Abadal}}, \bibinfo {author}
  {\bibfnamefont {D.}~\bibnamefont {Adler}}, \bibinfo {author} {\bibfnamefont
  {P.}~\bibnamefont {Weckesser}}, \bibinfo {author} {\bibfnamefont
  {A.}~\bibnamefont {Kruckenhauser}}, \bibinfo {author} {\bibfnamefont
  {V.}~\bibnamefont {Walther}}, \bibinfo {author} {\bibfnamefont
  {R.}~\bibnamefont {van Bijnen}}, \bibinfo {author} {\bibfnamefont
  {J.}~\bibnamefont {Rui}}, \bibinfo {author} {\bibfnamefont {C.}~\bibnamefont
  {Gross}}, \bibinfo {author} {\bibfnamefont {I.}~\bibnamefont {Bloch}},\ and\
  \bibinfo {author} {\bibfnamefont {J.}~\bibnamefont {Zeiher}},\ }\bibfield
  {title} {\bibinfo {title} {Realizing distance-selective interactions in a
  rydberg-dressed atom array},\ }\href
  {https://doi.org/10.1103/PhysRevLett.128.113602} {\bibfield  {journal}
  {\bibinfo  {journal} {Phys. Rev. Lett.}\ }\textbf {\bibinfo {volume} {128}},\
  \bibinfo {pages} {113602} (\bibinfo {year} {2022})}\BibitemShut {NoStop}%
\bibitem [{\citenamefont {Scholl}\ \emph {et~al.}(2022)\citenamefont {Scholl},
  \citenamefont {Williams}, \citenamefont {Bornet}, \citenamefont {Wallner},
  \citenamefont {Barredo}, \citenamefont {Henriet}, \citenamefont {Signoles},
  \citenamefont {Hainaut}, \citenamefont {Franz}, \citenamefont {Geier},
  \citenamefont {Tebben}, \citenamefont {Salzinger}, \citenamefont {Z\"urn},
  \citenamefont {Lahaye}, \citenamefont {Weidem\"uller},\ and\ \citenamefont
  {Browaeys}}]{PRXQuantum.3.020303}%
  \BibitemOpen
  \bibfield  {author} {\bibinfo {author} {\bibfnamefont {P.}~\bibnamefont
  {Scholl}}, \bibinfo {author} {\bibfnamefont {H.~J.}\ \bibnamefont
  {Williams}}, \bibinfo {author} {\bibfnamefont {G.}~\bibnamefont {Bornet}},
  \bibinfo {author} {\bibfnamefont {F.}~\bibnamefont {Wallner}}, \bibinfo
  {author} {\bibfnamefont {D.}~\bibnamefont {Barredo}}, \bibinfo {author}
  {\bibfnamefont {L.}~\bibnamefont {Henriet}}, \bibinfo {author} {\bibfnamefont
  {A.}~\bibnamefont {Signoles}}, \bibinfo {author} {\bibfnamefont
  {C.}~\bibnamefont {Hainaut}}, \bibinfo {author} {\bibfnamefont
  {T.}~\bibnamefont {Franz}}, \bibinfo {author} {\bibfnamefont
  {S.}~\bibnamefont {Geier}}, \bibinfo {author} {\bibfnamefont
  {A.}~\bibnamefont {Tebben}}, \bibinfo {author} {\bibfnamefont
  {A.}~\bibnamefont {Salzinger}}, \bibinfo {author} {\bibfnamefont
  {G.}~\bibnamefont {Z\"urn}}, \bibinfo {author} {\bibfnamefont
  {T.}~\bibnamefont {Lahaye}}, \bibinfo {author} {\bibfnamefont
  {M.}~\bibnamefont {Weidem\"uller}},\ and\ \bibinfo {author} {\bibfnamefont
  {A.}~\bibnamefont {Browaeys}},\ }\bibfield  {title} {\bibinfo {title}
  {{Microwave engineering of programmable $XXZ$ Hamiltonians in arrays of
  Rydberg atoms}},\ }\href {https://doi.org/10.1103/PRXQuantum.3.020303}
  {\bibfield  {journal} {\bibinfo  {journal} {PRX Quantum}\ }\textbf {\bibinfo
  {volume} {3}},\ \bibinfo {pages} {020303} (\bibinfo {year}
  {2022})}\BibitemShut {NoStop}%
\bibitem [{\citenamefont {W{\"u}ster}\ \emph {et~al.}(2011)\citenamefont
  {W{\"u}ster}, \citenamefont {Ates}, \citenamefont {Eisfeld},\ and\
  \citenamefont {Rost}}]{wuster2011excitation}%
  \BibitemOpen
  \bibfield  {author} {\bibinfo {author} {\bibfnamefont {S.}~\bibnamefont
  {W{\"u}ster}}, \bibinfo {author} {\bibfnamefont {C.}~\bibnamefont {Ates}},
  \bibinfo {author} {\bibfnamefont {A.}~\bibnamefont {Eisfeld}},\ and\ \bibinfo
  {author} {\bibfnamefont {J.}~\bibnamefont {Rost}},\ }\bibfield  {title}
  {\bibinfo {title} {{Excitation transport through Rydberg dressing}},\ }\href
  {https://doi.org/10.1088/1367-2630/13/7/073044} {\bibfield  {journal}
  {\bibinfo  {journal} {New J. Phys.}\ }\textbf {\bibinfo {volume} {13}},\
  \bibinfo {pages} {073044} (\bibinfo {year} {2011})}\BibitemShut {NoStop}%
\bibitem [{\citenamefont {Pagano}\ \emph {et~al.}(2024)\citenamefont {Pagano},
  \citenamefont {Jaschke}, \citenamefont {Weiss},\ and\ \citenamefont
  {Montangero}}]{PhysRevResearch.6.033282}%
  \BibitemOpen
  \bibfield  {author} {\bibinfo {author} {\bibfnamefont {A.}~\bibnamefont
  {Pagano}}, \bibinfo {author} {\bibfnamefont {D.}~\bibnamefont {Jaschke}},
  \bibinfo {author} {\bibfnamefont {W.}~\bibnamefont {Weiss}},\ and\ \bibinfo
  {author} {\bibfnamefont {S.}~\bibnamefont {Montangero}},\ }\bibfield  {title}
  {\bibinfo {title} {Optimal control transport of neutral atoms in optical
  tweezers at finite temperature},\ }\href
  {https://doi.org/10.1103/PhysRevResearch.6.033282} {\bibfield  {journal}
  {\bibinfo  {journal} {Phys. Rev. Res.}\ }\textbf {\bibinfo {volume} {6}},\
  \bibinfo {pages} {033282} (\bibinfo {year} {2024})}\BibitemShut {NoStop}%
\bibitem [{\citenamefont {Labuhn}\ \emph {et~al.}(2016)\citenamefont {Labuhn},
  \citenamefont {Barredo}, \citenamefont {Ravets}, \citenamefont
  {de~L{\'e}s{\'e}leuc}, \citenamefont {Macr{\`\i}}, \citenamefont {Lahaye},\
  and\ \citenamefont {Browaeys}}]{Labuhn2016}%
  \BibitemOpen
  \bibfield  {author} {\bibinfo {author} {\bibfnamefont {H.}~\bibnamefont
  {Labuhn}}, \bibinfo {author} {\bibfnamefont {D.}~\bibnamefont {Barredo}},
  \bibinfo {author} {\bibfnamefont {S.}~\bibnamefont {Ravets}}, \bibinfo
  {author} {\bibfnamefont {S.}~\bibnamefont {de~L{\'e}s{\'e}leuc}}, \bibinfo
  {author} {\bibfnamefont {T.}~\bibnamefont {Macr{\`\i}}}, \bibinfo {author}
  {\bibfnamefont {T.}~\bibnamefont {Lahaye}},\ and\ \bibinfo {author}
  {\bibfnamefont {A.}~\bibnamefont {Browaeys}},\ }\bibfield  {title} {\bibinfo
  {title} {Tunable two-dimensional arrays of single rydberg atoms for realizing
  quantum ising models},\ }\href {https://doi.org/10.1038/nature18274}
  {\bibfield  {journal} {\bibinfo  {journal} {Nature}\ }\textbf {\bibinfo
  {volume} {534}},\ \bibinfo {pages} {667} (\bibinfo {year}
  {2016})}\BibitemShut {NoStop}%
\bibitem [{\citenamefont {Scholl}\ \emph {et~al.}(2023)\citenamefont {Scholl},
  \citenamefont {Shaw}, \citenamefont {Tsai}, \citenamefont {Finkelstein},
  \citenamefont {Choi},\ and\ \citenamefont {Endres}}]{Scholl2023}%
  \BibitemOpen
  \bibfield  {author} {\bibinfo {author} {\bibfnamefont {P.}~\bibnamefont
  {Scholl}}, \bibinfo {author} {\bibfnamefont {A.~L.}\ \bibnamefont {Shaw}},
  \bibinfo {author} {\bibfnamefont {R.~B.-S.}\ \bibnamefont {Tsai}}, \bibinfo
  {author} {\bibfnamefont {R.}~\bibnamefont {Finkelstein}}, \bibinfo {author}
  {\bibfnamefont {J.}~\bibnamefont {Choi}},\ and\ \bibinfo {author}
  {\bibfnamefont {M.}~\bibnamefont {Endres}},\ }\bibfield  {title} {\bibinfo
  {title} {Erasure conversion in a high-fidelity rydberg quantum simulator},\
  }\href {https://doi.org/10.1038/s41586-023-06516-4} {\bibfield  {journal}
  {\bibinfo  {journal} {Nature}\ }\textbf {\bibinfo {volume} {622}},\ \bibinfo
  {pages} {273} (\bibinfo {year} {2023})}\BibitemShut {NoStop}%
\bibitem [{Note2()}]{Note2}%
  \BibitemOpen
  \bibinfo {note} {Views and opinions expressed are however those of the
  authors only and do not necessarily reflect those of the European Union or
  the European Research Council Executive Agency. Neither the European Union
  nor the granting authority can be held responsible for them.}\BibitemShut
  {Stop}%
\bibitem [{\citenamefont {Steinerberger}(2019)}]{Stefan2019}%
  \BibitemOpen
  \bibfield  {author} {\bibinfo {author} {\bibfnamefont {S.}~\bibnamefont
  {Steinerberger}},\ }\href {https://arxiv.org/abs/1902.00441} {\bibinfo
  {title} {A nonlocal functional promoting low-discrepancy point sets}}
  (\bibinfo {year} {2019}),\ \Eprint {https://arxiv.org/abs/1902.00441}
  {arXiv:1902.00441 [math.OC]} \BibitemShut {NoStop}%
\bibitem [{\citenamefont {Niederreiter}\ and\ \citenamefont
  {Wills}(1975)}]{Heiner_1975}%
  \BibitemOpen
  \bibfield  {author} {\bibinfo {author} {\bibfnamefont {H.}~\bibnamefont
  {Niederreiter}}\ and\ \bibinfo {author} {\bibfnamefont {J.~M.}\ \bibnamefont
  {Wills}},\ }\bibfield  {title} {\bibinfo {title} {Diskrepanz und distanz von
  ma{\ss}en bez{\"u}glich konvexer und jordanscher mengen},\ }\href
  {https://doi.org/10.1007/BF01190941} {\bibfield  {journal} {\bibinfo
  {journal} {Mathematische Zeitschrift}\ }\textbf {\bibinfo {volume} {144}},\
  \bibinfo {pages} {125} (\bibinfo {year} {1975})}\BibitemShut {NoStop}%
\bibitem [{\citenamefont {Niederreiter}(1992)}]{Harald1992}%
  \BibitemOpen
  \bibfield  {author} {\bibinfo {author} {\bibfnamefont {H.}~\bibnamefont
  {Niederreiter}},\ }\href {https://doi.org/10.1137/1.9781611970081} {\emph
  {\bibinfo {title} {Random Number Generation and Quasi-Monte Carlo Methods}}}\
  (\bibinfo  {publisher} {Society for Industrial and Applied Mathematics},\
  \bibinfo {year} {1992})\ \Eprint
  {https://arxiv.org/abs/https://epubs.siam.org/doi/pdf/10.1137/1.9781611970081}
  {https://epubs.siam.org/doi/pdf/10.1137/1.9781611970081} \BibitemShut
  {NoStop}%
\end{thebibliography}%

\clearpage
\onecolumngrid
\begin{center}
\textbf{Supplemental Material for}\\[3pt]``Emergent Causality of Macroscopic Transport and Equilibration in Long-Range Quantum Systems"
\end{center}
\setcounter{section}{0}
\setcounter{equation}{0}
\renewcommand{\thesection}{S\arabic{section}}
\renewcommand{\theequation}{S\arabic{equation}}
\renewcommand{\theHequation}{supp.\arabic{equation}}
\setcounter{secnumdepth}{2}

\tableofcontents{}
\section{Proof of Theorem 1}\label{sm:transport}
In this section, we give a detailed derivation of Theorem 1 in the main text. Let $\Lambda$ be the set of all the sites on a finite $D$-dimensional lattice, with the Euclidean distance (or periodic Euclidean distance on a torus). The lattice satisfies
\begin{equation}\label{origsm:eq: volume_grow}
    \sup_{x\in\Lambda}|x[r]\backslash x[r-1]|\leq \varphi r^{D-1},\qquad r\in\mathbb N,
\end{equation}
where $x[r]$ denotes a ball centered at the site $x$ with a radius $r$ and $|Z|$ represents the number of sites in the subset $Z\subseteq\Lambda$. We first consider a time-independent Hamiltonian on the lattice $\Lambda$ taking the form of
\begin{equation}\label{origsm:eq: Hamiltonian}
    H := \sum_{x \neq y \in \Lambda} J_{x y}  a_x^{\dagger} a_y + \sum_{Z \subseteq \Lambda} h_Z,
\end{equation}
where ${a_x}$ and $a_x^{\dagger}$ represent the bosonic annihilation and creation operators on the site $x \in \Lambda $ which satisfy the commutation relation $[a_x,a_y^{\dagger}]=\delta_{xy}$, and $h_Z$ denotes the interaction term on the subset $Z$, which is a self-adjoint function of local number operators $n_x:=a_x^{\dagger}a_x$. The hopping amplitude $J_{xy}$ obeys 
\begin{equation}
    J_{xy}=J_{yx}^{*},\ |J_{yx}|\leq\frac{J_0}{[1+|x-y|]^{D+\gamma}},
\end{equation}
where $J_0>0$ and we introduce $\gamma:=\alpha-D>0$ for convenience. Since the Hamiltonian satisfies the U(1) symmetry: $[H,N_{\Lambda}]=0$, where $N_{\Lambda}:=\sum_{x\in\Lambda}n_x$, we restrict our discussion to the sector $N_{\Lambda}=N\geq1$. This sector in the Hilbert space is finite-dimensional, and all the operations below are well defined even for unbounded occupation-dependent interactions on the full Fock space.

We consider the probability of transporting a macroscopic number of particles from the region $Y$ to the region $X$ within a time $t$. The two regions should be well separated from each other, with a distance $d_{XY}$ between the two regions defined as
\begin{equation}
    d_{XY}:=\min_{x\in X,y\in Y}|x-y|\geq1.
\end{equation}
Below, we omit the subscript $XY$ for simplicity. For $0\leq\theta_{\mathrm i}<\theta_{\mathrm f}\leq1$, set $\Delta\theta:=\theta_{\mathrm f}-\theta_{\mathrm i}\in(0,1]$. We aim to prove the inequality
\begin{equation}\label{eq: bound_center}
      \|P_{\bar{N}_X\geq\theta_{\mathrm f}}e^{-\mathrm{i}Ht}P_{\bar{N}_Y\geq1-\theta_{\mathrm i}}\|\leq\left(\frac{C_{\gamma}J_0|t|}{d^{\min(\gamma,1)}}\right)^{\Delta\theta N/2},
\end{equation}
with $d^{\min(\gamma,1)}$ replaced by $d/\ln(2+d)$ when $\gamma=1$. Here $P_{\bar{N}_X\geq\theta_{\mathrm f}}$ is a projection operator that projects the state on the Hilbert space $\mathcal{H}_{\bar{N}_X\geq\theta_{\mathrm f}}:=\text{span}\{|\bm{n}\rangle;\langle\bm{n}|N_X|\bm{n}\rangle\geq\theta_{\mathrm f} N\}$, and the norm of an operator $O$ is defined as $\|O\|=\|O\|_{\infty}:=\sup_{\|\psi\|=1}\|O|\psi\rangle\|$. Since we only consider one sector of the Hilbert space, the norm is also well defined. At $t=0$, the two number constraints have disjoint ranges, so the asserted bound holds; below we assume $t\ne0$.

We begin by specifying the relevant lattice geometry. The distance between a site $z$ and the subset $Z$ is defined as $d_Z(z)=\inf_{w\in Z}|z-w|$. Then we introduce a function $s:\Lambda\to\mathbb R$ by
\begin{equation}
    s(z)=d_Y(z)-d/2.
\end{equation}
The function $s(x)$ obeys the relation
\begin{equation}
    |s(z)-s(w)|\leq|z-w|,
\end{equation}
implying that the function is Lipschitz continuous. For $x\in X$ ($y\in Y$), we have $s(x)\geq d/2$ ($s(y)\leq -d/2$).

Next, we introduce a function
\begin{equation}
  F_x = f (2 s (x) / d).
\end{equation}
We require that $f (y)$ is a continuous function of $y$ with
$F_x = 1$ if $x \in X$ and $F_x = - 1$ if $x \in Y$. In this work, the function $f (x)$ is taken to be
\begin{equation}
f(x)=
\begin{cases}
1, & x \geqslant 1,\\
x, & -1 \leqslant x \leqslant 1,\\
-1, & x \leqslant -1.
\end{cases}
\end{equation}
For an arbitrary pair of points $(x, y)$, the function $F_x$ satisfies
\begin{equation}
  | F_x - F_y | \leqslant \min \left( 2, 2 \frac{|s (x) - s (y)|}{d} \right)
  \leqslant \min \left( 2, \frac{2 | x - y |}{d} \right),
\end{equation}
since the function $s (x)$ is Lipschitz continuous. Then we can define a geometric tilting operator $T_{\mu}$ as
\begin{equation}
  T_{\mu} = \exp [\mu A], \  A : = \sum_{x \in \Lambda} F_x n_x .
\end{equation}
Here $\mu > 0$ is a parameter to be determined later. Inserting the geometric tilting operator $T_{\mu}$ into the left-hand side of Eq. \eqref{eq: bound_center}, we obtain
\begin{equation}
  \| P_{\bar{N}_X \geqslant \theta_{\mathrm f}} T_{\mu}^{-1}T_{\mu} e^{- \mathrm{i} H t} T_{\mu}^{- 1}T_{\mu}
  P_{\bar{N}_Y \geqslant 1 - \theta_{\mathrm i}} \|=\| P_{\bar{N}_X \geqslant \theta_{\mathrm f}} T_{\mu}^{-1} e^{- \mathrm{i} \tilde{H} t} T_{\mu}
  P_{\bar{N}_Y \geqslant 1 - \theta_{\mathrm i}} \| ,
\end{equation}
where $\tilde{H}:=T_{\mu}HT_{\mu}^{-1}$. Following the same analysis as in Ref. \cite{Faupin2025}, we find that
\begin{align}
    \| P_{\bar{N}_X \geqslant \theta_{\mathrm f}} T_{\mu}^{- 1} \|&=\max_{\|\psi\|=1}\langle\psi|P_{\bar{N}_X \geqslant \theta_{\mathrm f}}\exp\left(-\mu\sum_xF_xn_x\right)|\psi\rangle\nonumber\\
    &=\max_{\|\psi\|=1}\langle\psi|P_{\bar{N}_X \geqslant \theta_{\mathrm f}}\exp\left(-\mu\sum_{x\in X}n_x+\mu\sum_{x\in Y}n_x-\mu\sum_{x\in\Lambda\backslash(X\bigcup Y)} F_xn_x\right)|\psi\rangle\nonumber\\
    &\leq\exp\left(-\mu\theta_{\mathrm f} N+\mu(1-\theta_{\mathrm f})N\right)\nonumber\\
    &=\exp(\mu N(1-2\theta_{\mathrm f})),\\
    \|  T_{\mu} P_{\bar{N}_Y\geqslant 1 - \theta_{\mathrm i}} \|&=\max_{\|\psi\|=1}\langle\psi| P_{\bar{N}_Y\geqslant 1 - \theta_{\mathrm i}}\exp\left(\mu\sum_xF_xn_x\right)|\psi\rangle\nonumber\\
    &=\max_{\|\psi\|=1}\langle\psi|P_{\bar{N}_Y\geqslant 1 - \theta_{\mathrm i}}\exp\left(\mu\sum_{x\in X}n_x-\mu\sum_{x\in Y}n_x+\mu\sum_{x\in\Lambda\backslash(X\bigcup Y)} F_xn_x\right)|\psi\rangle\nonumber\\
    &\leq \exp\left(\mu\theta_{\mathrm i} N-\mu(1-\theta_{\mathrm i})N\right)\nonumber\\
    &\leq \exp(\mu N(2\theta_{\mathrm i}-1)).
\end{align}
Therefore, we obtain
\begin{equation}
  \| P_{\bar{N}_X \geqslant \theta_{\mathrm f}} T_{\mu}^{- 1} \| \|  T_{\mu} P_{\bar{N}_Y
  \geqslant 1 - \theta_{\mathrm i}} \|  \leqslant e^{- 2 \mu (\theta_{\mathrm f} - \theta_{\mathrm i}) N}
  = e^{- 2 \mu \Delta\theta N} . \label{eq:PNX}
\end{equation}

Next, we need to appropriately choose the value of $\mu$ and further evaluate the upper bound of $\| T_{\mu} e^{- \mathrm{i} t H} T_{\mu}^{- 1} \|$. Writing $V:=\sum_Z h_Z$ and $J_{xx}:=0$, notice that $[T_{\mu}, V] = 0$ since $[n_x, V] =
0$ and
\begin{equation}
  T_{\mu} a_x^{\dagger} a_y T_{\mu}^{- 1} = e^{\mu (F_x - F_y)} a_x^{\dagger} a_y,
\end{equation}
the full Hamiltonian after the transformation is given by
\begin{equation}
  \tilde{H} = T_{\mu} H T_{\mu}^{- 1} = V + \sum_{x, y} J_{x y} e^{\mu (F_x - F_y)}
  a_x^{\dagger} a_y . \label{eq:tildeH}
\end{equation}
From the expression of Eq. \eqref{eq:tildeH}, we find that $\tilde{H}$ is generally not a Hermitian operator and the terms odd in $\mu(F_x-F_y)$ determine the anti-Hermitian part. The imaginary part of the operator $\tilde{H}$ takes the form of
\begin{equation}
  \text{Im} [\tilde{H}]:=\frac{\tilde H-\tilde H^\dagger}{2\mathrm{i}} = -\mathrm{i}\sum_{k = \text{odd}} \sum_{x, y} \frac{(\mu (F_x -
  F_y))^k}{k!} J_{x y} a_x^{\dagger} a_y = -\mathrm{i}\sum_{x, y} \sinh (\mu (F_x - F_y))
  J_{x y} a_x^{\dagger} a_y .
\end{equation}
Then we can upper bound the norm of $e^{- \mathrm{i} t \tilde{H}}$ as
\begin{equation}
  \| e^{- \mathrm{i} t \tilde{H}} \| \leqslant e^{| t | \| \text{Im} [\tilde{H}] \|},
\end{equation}
where $\|\text{Im}[\tilde H]\|$ can be upper bounded as
\begin{equation}
  \| \text{Im} [\tilde{H}] \| \leqslant \sup_{\|\psi\|=1}\sum_{x, y}|J_{xy}|\, | \sinh (\mu (F_x - F_y))
  |\,|\langle\psi|a_x^{\dagger}a_y|\psi\rangle| . \label{eq:ImH}
\end{equation}
By utilizing the inequality
\begin{equation}
  |\langle\psi|a_x^{\dagger} a_y|\psi\rangle|
  \leqslant \|a_x\psi\|\,\|a_y\psi\|
  \leqslant \frac{\langle\psi|(n_x+n_y)|\psi\rangle}{2},
\end{equation}
and the fact that $|J_{x y}| = |J_{y x}|$, we rewrite Eq. \eqref{eq:ImH}
as
\[
 \| \text{Im} [\tilde{H}] \| \leqslant \sup_{\|\psi\|=1}\sum_x\langle\psi|n_x|\psi\rangle\sup_x \sum_y |J_{x y}| |
   \sinh (\mu (F_x - F_y)) | = N \sup_x \sum_y |J_{x y}| | \sinh (\mu (F_x -
   F_y)) |.
\]
The same quadratic-form estimate and conjugation identity hold for fermionic and hard-core bosonic operators, so the argument applies to them as well.

Since
\begin{equation}
  | F_x - F_y | \leqslant \min \left\{ 2 \frac{| x - y |}{d}, 2 \right\},
\end{equation}
we introduce a length scale $R$ as
\begin{equation}
  R := \frac{d}{2 \mu} .
\end{equation}
Splitting the sum into the two regimes: $| x - y | \leqslant R$ and $| x - y
| > R$, we can rewrite \eqref{eq:ImH} as
\begin{eqnarray}
  \| \text{Im} [\tilde{H}] \| & \leqslant & \sup_x \sum_{y : | x - y |
  \leqslant R} |J_{x y}| N | \sinh (\mu (F_x - F_y)) | \nonumber\\
  &  & + \sup_x \sum_{y : | x - y | > R} |J_{x y}| N | \sinh (\mu (F_x - F_y))
  | . \label{eq:ImH2} 
\end{eqnarray}
To evaluate $\| \text{Im} [\tilde{H}] \|$, we first consider
\begin{equation}
  \sinh (x) := \sum_{k = \text{odd}} \frac{x^k}{k!} = x + \frac{x^3}{3!}
  + \cdots
\end{equation}
For $| x | > 1$, we use the relation $|\sinh(x)|\leqslant e^{|x|}$. On the
other hand, for $|x|\leq1$, we have
\begin{equation}
  |\sinh x|=\sum_{j = 0}^{\infty} \frac{|x|^{2 j + 1}}{(2 j + 1) !} \leqslant \sum_{j =
  0}^{\infty} \frac{1}{(2 j + 1) !} | x | = \frac{e^2 - 1}{2 e} | x |,
\end{equation}
which gives rise to the following bound when $|x-y|\leqslant R$:
\begin{equation}
  | \sinh (\mu (F_x - F_y)) | \leqslant \frac{e^2 - 1}{2 e} \frac{2 \mu | x -
  y |}{d} .
\end{equation}
Therefore, the right-hand side of Eq. \eqref{eq:ImH2} can be rewritten as
\[ \| \text{Im} [\tilde{H}] \| \leqslant \sup_x \sum_{y : | x - y | \leqslant
   R} |J_{xy}| \frac{e^2 - 1}{2 e} N 2 \mu \frac{| x - y |}{d} + \sup_x \sum_{y
   : | x - y | > R} |J_{xy}| N e^{2 \mu} . \]
Since we assume
\begin{equation}
  |J_{xy}| \leqslant \frac{J_0}{[1 + | x - y |]^{D + \gamma}},
\end{equation}
we obtain
\begin{equation}
  \| \text{Im} [\tilde{H}] \| \leqslant J_0 \frac{e^2 - 1}{2 e} N \frac{2
  \mu}{d} \kappa_1 (R) + J_0 N e^{2 \mu} \tau (R),
\end{equation}
where we define
\begin{eqnarray*}
  \kappa_1 (R) & := & \sup_x \sum_{y : | x - y | \leqslant R} \frac{1}{[1
  + | x - y |]^{D + \gamma}} | x - y |,\\
  \tau (R) & = & \sup_x \sum_{y : | x - y | > R} \frac{1}{[1 + | x - y |]^{D +
  \gamma}} .
\end{eqnarray*}
For $0<R<1$, the volume-growth bound gives $\kappa_1(R)\leqslant CR$ and $\tau(R)\leqslant C\leqslant CR^{-\gamma}$. Therefore, the estimates below also apply in this range:
\begin{equation}
  \| e^{- \mathrm{i} t \tilde{H}} \| \leqslant e^{| t | \| \text{Im} [\tilde{H}] \|}
  \leqslant \exp \left[ \frac{e^2 - 1}{2 e} J_0 N \left( \frac{2 \mu}{d}
  \kappa_1 (R) + e^{2 \mu} \tau (R) \right) | t | \right] .
\end{equation}

When $0 <
\gamma < 1$, we have
\begin{eqnarray}
  \kappa_1 (R) & \leqslant & \varphi_1 R^{1 - \gamma}, \label{eq:kappa} \\
  \tau (R) & \leqslant & \varphi' R^{- \gamma}, 
\end{eqnarray}
where $\varphi_1, \varphi' > 0$ are constants depending on the structure of
the lattice and the decay exponent $\gamma$. Hence,
\begin{eqnarray}
  \| e^{- \mathrm{i} t \tilde{H}} \| & \leqslant & \exp \left[ \frac{e^2 - 1}{2 e}
  (\varphi_1 + \varphi' e^{2 \mu}) J_0 N R^{- \gamma} | t | \right]
  \nonumber\\
  & \leqslant & \exp \left[ \frac{e^2 - 1}{2 e} \max (\varphi_1, \varphi') (1
  + e^{2 \mu}) J_0 N R^{- \gamma} | t | \right] \label{eq:49} 
\end{eqnarray}
By choosing
\begin{equation}
  e^{2 \mu} (2 \mu)^{\gamma} = \frac{d^{\gamma}}{C_1 J_0 | t |},
  \label{eq:choicemu}
\end{equation}
where
\begin{equation}
  C_1 := \frac{e^2 - 1}{2 e} \max (\varphi_1, \varphi') > 0,
\end{equation}
we have
\begin{equation}
  \| e^{- \mathrm{i} t \tilde{H}} \| \leqslant \exp [N (1 + e^{- 2 \mu})] \leqslant
  e^{2 N} .
\end{equation}
Here we only consider the case: $d^{\gamma} / C_1 J_0 | t | > 1$, which is
outside the light cone and constraints $\mu$ to be positive. When $d^{\gamma}
/ C_1 J_0 | t |$ is large, we can approximate $\mu$ as
\begin{equation}
  \mu = \frac{1}{2} \ln \left( \frac{d^{\gamma}}{C_1 J_0 | t |} \right) - O
  \left( \text{lnln} \left( \frac{d^{\gamma}}{C_1 J_0 | t |} \right) \right) .
\end{equation}
On the other hand, we have
\begin{equation}
  2 \mu \geqslant \frac{1}{1 + \gamma} \ln \left( \frac{d^{\gamma}}{C_1 J_0 |
  t |} \right),
\end{equation}
since
\begin{equation}
  \ln \left( \frac{d^{\gamma}}{C_1 J_0 | t |} \right) = 2 \mu + \gamma \ln (2
  \mu) \leqslant (1 + \gamma) 2 \mu .
\end{equation}

Hence, combining Eqs. \eqref{eq:PNX} and \eqref{eq:49}, and restricting to $C_1J_0|t|e^{2(1+\gamma)/(\Delta\theta)}/d^\gamma<1$, we obtain
\begin{eqnarray}
  \| P_{\bar{N}_X \geqslant \theta_{\mathrm f}} T_{\mu}^{- 1} e^{- \mathrm{i} t \tilde{H}} T_{\mu}
  P_{\bar{N}_Y \geqslant 1 - \theta_{\mathrm i}} \| & \leqslant & \exp [- 2 \mu \Delta\theta N +
  2 N] \nonumber\\
  & \leqslant & \left( \frac{C_1 J_0 | t |}{d^{\gamma}} e^{2 \frac{1 +
  \gamma}{\Delta\theta}} \right)^{\Delta\theta N / (1 + \gamma)} \nonumber\\
  & \leqslant & \left( \frac{C_1' J_0 | t |}{d^{\gamma}} \right)^{\Delta\theta N /
  2}, \label{eq:MASST1} 
\end{eqnarray}
where $C_1' := C_1 \times \exp [2 (1 + \gamma) /(\Delta\theta)] > C_1$. Hence,
we need to revise the light cone structure as
\begin{equation}
  d^{\gamma} = C_1' J_0 | t | .
\end{equation}
When $d^{\gamma} / C_1' J_0 | t | > 1$ is satisfied, the previous condition
$d^{\gamma} / C_1 J_0 | t | > 1$ is automatically satisfied as well. If $C_1'J_0|t|/d^\gamma\geq1$, Eq.~\eqref{eq:MASST1} follows from unitarity.

For the case $\gamma = 1$, the estimate for $\tau (R)$ remains
invariant, whereas Eq. \eqref{eq:kappa} should be modified as
\begin{equation}
  \kappa_1 (R) \leqslant \varphi_c \ln(2+R),
\end{equation}
where the logarithmic growth originates from the limit of $(R^{1-\gamma}-1)/(1-\gamma)$ as $\gamma\to1$. Performing a
similar analysis, we obtain an upper bound of $\| e^{- \mathrm{i} t \tilde{H}} \|$ as
\begin{eqnarray}
  \| e^{- \mathrm{i} t \tilde{H}} \| & \leqslant & \exp \left[ \frac{e^2 - 1}{2 e}
  (\varphi_c \ln(2+R) + \varphi' e^{2 \mu}) J_0 N R^{- 1} | t | \right]
  \nonumber\\
  & \leqslant & \exp \left[ C_c (\ln(2+R) + e^{2 \mu}) J_0 N \frac{2 \mu}{d} | t
  | \right] . 
\end{eqnarray}
Here we define
\begin{equation}
  C_c := \frac{e^2 - 1}{2 e} \max (\varphi_c, \varphi') .
\end{equation}
However, a similar choice of $\mu$ as in Eq. \eqref{eq:choicemu} cannot fully
eliminate the logarithmic correction: $\ln(2+d / 2 \mu)$. Here we instead choose
\begin{equation}
  2 \mu e^{2 \mu} = \frac{d}{C_c J_0 | t | \ln(2+d)} .
\end{equation}
By requiring $2 \mu \geqslant 1$, we have $R\leq d$ and therefore
\begin{eqnarray}
  \| e^{- \mathrm{i} t \tilde{H}} \| & \leqslant & \exp \left[ N (\ln(2+R) + 1)
  \frac{1}{\ln(2+d)} \right] \nonumber\\
  & \leqslant & e^{2 N} . \label{eq:bound2} 
\end{eqnarray}
The condition $2 \mu \geqslant 1$ is equivalent to $d / [C_c J_0 | t | \ln(2+d)]
\geqslant e$. Similarly, we also have
\begin{equation}
  2 \mu \geqslant \frac{1}{2} \ln \left( \frac{d}{C_c J_0 | t | \ln(2+d)} \right)
  .
\end{equation}
Combining Eq. \eqref{eq:bound2} with Eq. \eqref{eq:PNX}, we obtain
\begin{eqnarray}
  \| P_{\bar{N}_X \geqslant \theta_{\mathrm f}} T_{\mu}^{- 1} e^{- \mathrm{i} t \tilde{H}} T_{\mu}
  P_{\bar{N}_Y \geqslant 1 - \theta_{\mathrm i}} \| & \leqslant & \exp [- 2 \mu \Delta\theta N +
  2 N] \nonumber\\
  & \leqslant & \left( e^{4 /(\Delta\theta)} \frac{C_c J_0 | t | \ln(2+d)}{d}
  \right)^{\Delta\theta N / 2} . \label{eq:MASST2} 
\end{eqnarray}
Since $e^{4 /(\Delta\theta)} > e^2$, this result identifies the effective light cone $d / \ln(2+d) = C_c'
J_0 | t |$, where $C_c' := C_c e^{4 /(\Delta\theta)}$. If $d / \ln(2+d)
\geqslant C_c' J_0 | t |$ is satisfied, the condition $2 \mu > 1$ is also
satisfied. If $C_c'J_0|t|\ln(2+d)/d\geq1$, Eq.~\eqref{eq:MASST2} follows from unitarity. Although the effective light cone acquires a logarithmic correction, its power-law exponent tends to one in the limit $d\rightarrow\infty$ since
\begin{equation}
 \frac{d}{\ln(2+d)}=d^{1-\ln(\ln(2+d))/\ln d}=d^{1-\epsilon},
\end{equation}
where $\epsilon:=\ln(\ln(2+d))/\ln d\rightarrow0$ vanishes as $d\rightarrow\infty$.

For the last case: $\gamma > 1$, we have
\begin{eqnarray}
  \kappa_1 (R) & \leqslant & \varphi_2, \\
  \tau (R) & \leqslant & \varphi_2' R^{- \gamma}, 
\end{eqnarray}
where $\varphi_2$ is an $\mathcal{O} (1)$ constant. Following a similar
procedure, we find that
\begin{eqnarray}
  \| e^{- \mathrm{i} t \tilde{H}} \| & \leqslant & \exp \left[ \frac{e^2 - 1}{2 e} \max
  (\varphi_2, \varphi_2') (R^{- 1} + e^{2 \mu} R^{- \gamma}) J_0 N | t |
  \right] \nonumber\\
  & =: & \exp [(R^{- 1} + e^{2 \mu} R^{- \gamma}) C_2 J_0 N | t |],
  \label{eq:58} 
\end{eqnarray}
where
\begin{equation}
  C_2 := \frac{e^2 - 1}{2 e} \max (\varphi_2, \varphi_2') .
\end{equation}
In this case, we cannot choose
\begin{equation}
  e^{2 \mu} (2 \mu)^{\gamma} = \frac{d^{\gamma}}{C_2 J_0 | t |},
\end{equation}
since the term proportional to $R^{- 1}$ may diverge. Instead, we choose
\begin{equation}
  e^{2 \mu} = \sqrt{\frac{d}{C_2 J_0 | t |}} .
\end{equation}
The first term in the exponent of Eq. \eqref{eq:58} can be thus given by
\begin{equation}
  \frac{2 \mu}{d} C_2 J_0 N | t | = \frac{1}{2} \ln \left( \frac{d}{C_2 J_0 |
  t |} \right) \frac{C_2 J_0 N | t |}{d} .
\end{equation}
By defining $x := d / (C_2 J_0 | t |) > 1$, we can further bound it as
\begin{equation}
  \frac{2 \mu}{d} C_2 J_0 N | t | = N \frac{\ln x}{2 x} \leqslant \frac{N}{2
  e} .
\end{equation}
The second term in the exponent of Eq. \eqref{eq:58} can be similarly bounded
by
\begin{eqnarray}
  e^{2 \mu} \left( \frac{2 \mu}{d} \right)^{\gamma} C_2 J_0 N | t | &
  = & N d^{1-\gamma}\left[ \frac{1}{2} \ln \left( \frac{d}{C_2 J_0 | t |} \right)
  \right]^{\gamma} \sqrt{\frac{C_2 J_0 | t |}{d}} \nonumber\\
  & \leqslant & N \frac{[\ln x]^{\gamma}}{2^{\gamma}\sqrt{x}} \nonumber\\
  & \leqslant & N \left( \frac{\gamma}{e} \right)^{\gamma} . 
\end{eqnarray}
Thus, Eq. \eqref{eq:58} yields the bound
\begin{equation}
  \| e^{- \mathrm{i} t \tilde{H}} \| \leqslant \exp \left[ N \left( \frac{1}{2 e} +
  \left( \frac{\gamma}{e} \right)^{\gamma} \right) \right] .
  \label{eq:bound3}
\end{equation}
By using Eq. \eqref{eq:bound3} and Eq. \eqref{eq:PNX}, we obtain
\begin{eqnarray}
  \| P_{\bar{N}_X \geqslant \theta_{\mathrm f}} T_{\mu}^{- 1} e^{- \mathrm{i} t \tilde{H}} T_{\mu}
  P_{\bar{N}_Y \geqslant 1 - \theta_{\mathrm i}} \| & \leqslant & \exp \left[ - 2 \mu
  \Delta\theta N + N \left( \frac{1}{2 e} + \left( \frac{\gamma}{e}
  \right)^{\gamma} \right) \right] \nonumber\\
  & = & \left( \frac{C_2' J_0 | t |}{d} \right)^{\Delta\theta N / 2},
  \label{eq:MASST3} 
\end{eqnarray}
where
\begin{equation}
  C_2' := C_2 \exp \left[ \frac{2}{\Delta\theta} \left( \frac{1}{2 e} + \left(
  \frac{\gamma}{e} \right)^{\gamma} \right) \right] > C_2 .
\end{equation}
If $C_2'J_0|t|/d\geq1$, Eq.~\eqref{eq:MASST3} also follows from unitarity. We emphasize that even though the prefactor $(\gamma / e)^{\gamma}$ seems to grow rapidly with $\gamma$, the usual value of $\gamma$ is not large in real
experiments. For the long-range hopping in a Rydberg atom array, the maximal
value of $\gamma$ is considered to be 5, which leads to the contribution
$(\gamma / e)^{\gamma} \simeq 21.06$.

Combining Eqs. \eqref{eq:MASST1}, \eqref{eq:MASST2}, and \eqref{eq:MASST3}, we
finally arrive at the bound
\begin{equation}
  \| P_{\bar{N}_X \geqslant \theta_{\mathrm f}} e^{- \mathrm{i} H t} P_{\bar{N}_Y \geqslant 1 -
  \theta_{\mathrm i}} \| \leqslant \left( \frac{C J_0 | t |}{d^{\gamma_{\varepsilon}}}
  \right)^{\Delta\theta N/2},
\end{equation}
where $C$ is a constant depending only on the geometric structure, decay exponent
$\gamma$ and transported fraction $\Delta\theta$, with $d^{\gamma_\varepsilon}$ replaced by $d/\ln(2+d)$ for
the case $\gamma = 1$, and
\begin{equation}
  \gamma_{\varepsilon} := \min (\gamma, 1) .
\end{equation}
The proof is thus completed. For locally integrable time-dependent hopping coefficients with the same uniform bound, replace $e^{-\mathrm{i}tH}$ by $\mathcal T\exp[-\mathrm{i}\int_0^tH(s)\,ds]$; differentiation of the tilted evolution gives the same estimate, so all bounds remain unchanged.

\section{Proof of Theorem 2}\label{sm:lower-proof}

We bound the equilibrium probability by the probability of occupying a distant region. For the region $B_j$ in Theorem 2 in the main text, the definition of $\mathcal H_{\mathrm{eq}}$ and $|B_j|\geqslant\eta L^D$ gives
\begin{equation}
 P_{\mathrm{eq}}\leqslant P_{N_j\geqslant(1-\epsilon)\eta N}.
\end{equation}
Apply Theorem 1 in the main text with $X=B_j$, $\theta_{\mathrm i}=0$ and $\theta_{\mathrm f}=(1-\epsilon)\eta$. Since $d_{B_j,Y}\geqslant\eta L$, for $\alpha\ne D+1$,
\begin{equation}
 \langle\Psi(t)|P_{\mathrm{eq}}|\Psi(t)\rangle
 \leqslant\left(\frac{CJ_0t}{L^{\min(\alpha-D,1)}}\right)^{(1-\epsilon)\eta N},
 \qquad t\geqslant0,
\end{equation}
with $L^{\min(\alpha-D,1)}$ replaced by $L/\ln(2+L)$ when $\alpha=D+1$. Here $C$ depends only on the fixed parameters, including $\epsilon,\eta$. For a sufficiently small fixed $c>0$, this implies $\langle\Psi(t)|P_{\mathrm{neq}}|\Psi(t)\rangle\geqslant1-e^{-c'N}$ throughout $0\leqslant t\leqslant cL^{\min(\alpha-D,1)}/J_0$, with the same critical replacement. Consequently, for every fixed $\delta\in(0,1)$ and all sufficiently large $L$,
\begin{equation}
 \frac1T\left|\mathcal T^{[0,T]}_{\delta,|\Psi(0)\rangle}\right|
 \geqslant\min\left\{1,\frac{cL^{\min(\alpha-D,1)}}{J_0T}\right\},
\end{equation}
again with the critical replacement. A sequence satisfying the equilibration criterion must therefore obey the lower bound in Theorem 2. Although the proof above assumes that $\eta>0$ is independent of the system size, the same lower bound extends to the single-site partition, corresponding to $\eta=1/|\Lambda|$. Indeed, we can choose $X=\bigcup_{j\in\mathcal J}B_j$ such that $|X|\geqslant c_V|\Lambda|$, with $\theta_f=(1-\epsilon)c_V$ and $d_{X,Y}\geqslant c_D L$, where $c_V,c_D>0$ are independent of $L$. The analysis remains invariant.

\section{Optimality of the Light Cone in Theorem 1}\label{sm:optimality}
According to Theorem 1 in the main text, for $\alpha\ne D+1$ the light-cone structure is given by $t\propto d^{\min(\alpha-D,1)}$. We construct a Hamiltonian such that $N$ bosons initially occupying the site $y$ can be transferred to the site $x\ne y$ within a time $t$ perfectly, i.e., $\langle \psi_x^N
| \mathcal{T} \left[ \exp[- \mathrm{i} \int_0^t d t' H (t')]
  \right] | \psi_y^N \rangle = 1$, where
\begin{equation}
  | \psi_y^N \rangle = \frac{(a_y^{\dagger})^N}{\sqrt{N!}} | \text{vac}
  \rangle, \quad | \psi_x^N \rangle = \frac{(a_x^{\dagger})^N}{\sqrt{N!}} |
  \text{vac} \rangle .
\end{equation}
All density-dependent interactions are set to zero. At $\alpha=D+1$, the construction below gives a linear transfer time and does not close the logarithmic gap in the lower bound.

For the case $\alpha \geqslant D + 1$, choose a nearest-neighbor path $z_0=y,z_1,\ldots,z_\ell=x$ on $\mathbb Z^D$, with $\ell=\sum_{\mu=1}^D|x^\mu-y^\mu|\leqslant\sqrt D\,|x-y|$. We consider the Hamiltonian
\begin{equation}
  H_j = \mathrm{i} \frac{J_0}{2^\alpha} (a_{z_{j+1}}^{\dagger} a_{z_j} - a_{z_j}^{\dagger} a_{z_{j+1}}),
\end{equation}
with $H(t)=H_j$ when $t \in \left[ j \frac{\pi2^{\alpha-1}}{J_0}, (j + 1) \frac{\pi2^{\alpha-1}}{J_0} \right)$
for $j=0,\ldots,\ell-1$, and $H(t) = 0$ elsewhere. Only this one bond is active during each step; its coefficient satisfies $|J_{z_{j+1},z_j}|=J_0/(1+|z_{j+1}-z_j|)^\alpha$. Then the bosonic operator
$a_{z_j}^{\dagger}$ evolves during a step of duration $u$ as
\begin{equation}
  e^{-\mathrm{i} H_j u} a_{z_j}^{\dagger} e^{\mathrm{i} H_j u} = \cos \left(\frac{J_0u}{2^\alpha}\right) a_{z_j}^{\dagger} + \sin \left(\frac{J_0u}{2^\alpha}\right) a_{z_{j+1}}^{\dagger} .
\end{equation}
Therefore,
\begin{equation}
  | \psi_y^N (t) \rangle = \mathcal{T} \left[ e^{- \mathrm{i} \int_0^t d t' H (t')}
  \right] \frac{(a_y^{\dagger})^N}{\sqrt{N!}} | \text{vac} \rangle .
\end{equation}
After the first period $j = 0$, the state is transferred into
\begin{equation}
  \left| \psi_y^N \left( \frac{\pi2^{\alpha-1}}{J_0} \right) \right\rangle =
  \frac{(a_{z_1}^{\dagger})^N}{\sqrt{N!}} | \text{vac} \rangle .
\end{equation}
Each step transfers all particles perfectly along the path in a time $\pi2^{\alpha-1}/J_0$. Sequentially applying these steps therefore achieves perfect state transfer at
\begin{equation}
 t = \frac{\pi2^{\alpha-1}\ell}{J_0}\leqslant\frac{\pi2^{\alpha-1}\sqrt D\,|x-y|}{J_0} .
\end{equation}

For the case $D < \alpha < D + 1$, we can still use the state-transfer scheme of Ref.~\cite{Tran_hierarchy_2020}. By translation, take $y=0$. We first
define $q \in \mathbb{Z}^+$ as
\begin{equation}
  q := \lfloor \log_2 |x| \rfloor + 1.
\end{equation}
Then let $B_q\subset\mathbb Z^D$ be a lattice cube containing $0$ and $x$, with $2^q$ sites on each side, and let $B_{q-1}^{(0)},\ldots,B_1^{(0)}$ be its dyadic subcubes containing $0$, so that
\begin{equation}
  \{ 0 \}=B_0^{(0)} \subset B_1^{(0)} \subset \cdots \subset B_{q - 1}^{(0)} \subset
  B_q^{(0)} = B_q .
\end{equation}
Similarly, we have
\begin{equation}
  \{ x \}=B_0^{(x)} \subset B_1^{(x)} \subset \cdots \subset B_{q - 1}^{(x)} \subset
  B_q^{(x)} = B_q .
\end{equation}
For $z\in\{0,x\}$ and $s=0,\ldots,q$, these cubes satisfy
\begin{equation}
 B_s^{(z)}\in\bigl\{b+\{0,\ldots,2^s-1\}^D:b\in\mathbb Z^D\bigr\},
 \qquad |B_s^{(z)}|=2^{sD}.
\end{equation}
They exist because $2^q>|x|$ permits a common cube $B_q$, and repeated dyadic subdivision gives the two chains.
The transfer protocol is
\begin{equation}
  \{ 0 \} \rightarrow B_1^{(0)} \rightarrow \cdots \rightarrow B_{q - 1}^{(0)}
  \rightarrow B_q \rightarrow B_{q - 1}^{(x)} \rightarrow \cdots \rightarrow
  B_1^{(x)} \rightarrow \{ x \},
\end{equation}
with $2q$ steps in total. The construction is supported on $B_q$, so a finite lattice must contain this cube. For an expansion step, take the disjoint regions $A=B_{s-1}^{(z)}$ and $B=B_s^{(z)}\setminus B_{s-1}^{(z)}$. For the state transfer from the region $A$ to the larger region $A\cup B$, we choose the Hamiltonian
\begin{equation}
  H_{AB} = \sum_{k \in A, j \in B} \mathrm{i} C (a_j^{\dagger} a_k - a_k^{\dagger} a_j) .
  \label{eq:HAB}
\end{equation}
By introducing the transformed operators
\begin{equation}
  b_A^{\dagger} := \sum_{i \in A} \frac{a_i^{\dagger}}{\sqrt{| A |}},
  \quad b_B^{\dagger} := \sum_{j \in B} \frac{a_j^{\dagger}}{\sqrt{| B |}},
\end{equation}
we can rewrite Eq. \eqref{eq:HAB} as
\begin{equation}
  H_{A B} = \mathrm{i} g (b_B^{\dagger} b_A - b_A^{\dagger} b_B),
\end{equation}
where $g = C \sqrt{| A | | B |}$. Since $A\cap B=\varnothing$, the operators $b_A$ and
$b_B$ satisfy the bosonic commutation relations
\begin{equation}
  [b_A,b_B]=[b_A,b_B^\dagger]=0, \quad [b_A, b_A^{\dagger}] = [b_B,b_B^{\dagger}] = 1.
\end{equation}
The evolution of the operator $b_A^{\dagger}$ is given by
\begin{equation}
  e^{-\mathrm{i} H_{AB}t} b_A^{\dagger} e^{\mathrm{i} H_{AB}t} = \cos (g t) b_A^{\dagger} + \sin (g t)
  b_B^{\dagger} .
\end{equation}
If the initial state in the region $A$ is given by $(b_A^{\dagger})^N |\text{vac}\rangle/\sqrt{N!}$, the transferred state at time
\begin{equation}
 t=\frac{\arccos\sqrt{|A|/(|A|+|B|)}}{C\sqrt{|A||B|}}
\end{equation}
is given by
\begin{equation}
 |\psi(t)\rangle=\frac{(b_{A\cup B}^{\dagger})^N}{\sqrt{N!}}|\text{vac}\rangle,
 \qquad b_{A\cup B}^{\dagger}=\frac{\sum_{j\in A\cup B}a_j^\dagger}{\sqrt{|A|+|B|}} .
\end{equation}

By construction, we choose to switch on the Hamiltonian $H_{A B}$ for the
state transferred from the region $A$ to the region $A\cup B$ and switch off all the
other Hamiltonians. For the contraction along the $x$-chain, we apply the negative of the corresponding Hamiltonians in reverse order, for the same durations. Then the total time for the perfect state transfer is
given by
\begin{equation}
  \tau = 2 \sum_{s = 1}^q T_{B, s},
\end{equation}
where $T_{B,s}$ is the time of state transfer from $B_{s-1}^{(0)}$ to $B_s^{(0)}$ or from $B_{s-1}^{(x)}$ to $B_s^{(x)}$. Every active edge at step $s$ has length at most $\sqrt D(2^s-1)$. To satisfy the prescribed power-law hopping bound, we choose
\begin{equation}
 C=C_s=\frac{J_0}{[1+\sqrt D(2^s-1)]^\alpha} .
\end{equation}
Since $|A|=2^{(s-1)D}$ and $|B|=(2^D-1)2^{(s-1)D}$, the time $T_{B,s}$ is upper bounded by
\begin{equation}
 T_{B,s}=\frac{\arccos(2^{-D/2})}{C_s2^{(s-1)D}\sqrt{2^D-1}}
 \leqslant\frac{C_{\alpha,D}\,2^{(\alpha-D)s}}{J_0},
\end{equation}
where here and below $C_{\alpha,D}$ depends only on $\alpha$ and $D$. Then the total time is upper bounded by
\begin{equation}
 \tau\leqslant\frac {C_{\alpha,D}}{J_0}\sum_{s=1}^q2^{(\alpha-D)s}
 \leqslant\frac{C_{\alpha,D}\,2^{q(\alpha-D)}}{J_0}
 \leqslant\frac{C_{\alpha,D}|x-y|^{\alpha-D}}{J_0} .
\end{equation}
The last inequality uses $2^q\leqslant2|x|$ for $y=0$ and $x\ne0$. Both protocols leave the vacuum fixed and map $a_y^\dagger$ to $a_x^\dagger$ under conjugation by $U(t,0)=\mathcal{T} \left[ \exp[- \mathrm{i} \int_0^t d t' H (t')]
  \right]$, so the transfer amplitude is one for every $N$, with a transfer time independent of $N$.

\section{Acceleration of macroscopic equilibration by long-range hopping}

In this section, we estimate the equilibration time for a long-range non-interacting fermion chain under the periodic boundary condition and show that the macroscopic equilibration can be intrinsically accelerated by the long-range hopping. The proof mainly follows the time-averaging approach in Ref. \cite{HaraKoike2026}, supplemented by an estimate of the distribution of single-particle energy differences.

\subsection{Model and the Main Result}
Let $\Lambda_L =\mathbb{Z}/L\mathbb{Z}$ where the system size $L$ is even. For $r \in \mathbb{Z}$ and $1 \leqslant r \leqslant L - 1$, we define $d_L (r) = \min (r, L - r)$. We consider the Hamiltonian
\begin{equation}
  H = \sum_{x \in \Lambda_L} \sum_{r = 1}^{L - 1} J_L (r) c_{x + r}^{\dagger}
  c_x,  \label{eq:Heq}
\end{equation}
where the hopping amplitude is given by
\begin{equation}
  J_L (r) = \frac{J_0}{d_L (r)^{\alpha}} + \frac{J_0}{L^{\alpha}} \cos \left(
  \frac{\pi r^2}{L} \right),
\end{equation}
with the exponent $1 < \alpha < 3 / 2$ and $J_0>0$. Since $L$ is even, the hopping
amplitude is symmetric around $r = L / 2$:
\begin{equation}
  J_L (L - r) = J_L (r) .
\end{equation}
The Hamiltonian $H$ is Hermitian and translation invariant. Moreover, since
$d_L (r) \leqslant L / 2$, the amplitude $J_L (r)$ satisfies
\begin{equation}
  (1 - 2^{- \alpha}) \frac{J_0}{d_L (r)^{\alpha}} \leqslant J_L (r) \leqslant
  (1 + 2^{- \alpha}) \frac{J_0}{d_L (r)^{\alpha}} .
\end{equation}
We work in the $N$-particle sector at fixed density $\bar{\rho}=N / L\in (0, 1)$. We divide the whole lattice into $n$ boxes with a box centered at $c$ defined as $B (c) : = \left\{ x \in \Lambda : |x - c| <l_c/2 \right\}$. Here we take $l_c = \kappa L$ where $\kappa$ is an $O$(1) constant. Then the number of the boxes is $n = \lceil L / l_c \rceil$. Define $n_x=c_x^{\dagger}c_x$ and
\begin{equation}
  \rho_c = \frac{1}{l_c} \sum_{x \in B (c)} n_x, \quad \Delta \rho_c = \rho_c
  - \bar{\rho} = \rho_c - \frac{N}{L} .
\end{equation}
Let $| e_j \rangle$ become the eigenvector of $\Delta \rho_c$: $\Delta \rho_c| e_j \rangle = \lambda_{j, c} | e_j \rangle$. For a tolerance $\varepsilon>0$, the equilibrium subspace is defined as
\begin{equation}
  \mathcal{H}_{\text{eq}} := \text{span} \left\{ | e_j \rangle : |
  \lambda_{j, c} | \leqslant \varepsilon \bar{\rho} \quad \text{for all $c$}\right\},
\end{equation}
and let the operator $P_{\text{eq}}$ denote the projector onto the equilibrium subspace $\mathcal{H}_{\text{eq}}$. Similarly, we can also define the nonequilibrium subspace and its projector $P_{\text{neq}} = I - P_{\text{eq}}$.

Our main result is as follows:
\begin{theorem}
For an initial state $|\Psi(0)\rangle$ in the $N$-fermion sector, let $|\Psi(t)\rangle=e^{-iHt}|\Psi(0)\rangle$ and define the set of nonequilibrium times during the time
interval $[0, \tau]$ as
\[ \mathcal{T}_{\delta, | \Psi (0) \rangle}^{[0, \tau]} := \{ t \in [0,
   \tau] : \langle \Psi (t) | P_{\text{neq}} | \Psi (t) \rangle > \delta \},
\]
and let $| \mathcal{T}_{\delta, | \Psi (0) \rangle}^{[0, \tau]} |$ denote its Lebesgue measure (the length of the set). Let $\tau_L = L^{\alpha - 1 / 2} g (L) / J_0$, where $g (L)$ is an
arbitrary function satisfying $\lim_{L \rightarrow \infty} g (L) = \infty$. Then for
every $T > \tau_L$ and every fixed $\delta>0$, we always have
\begin{equation}
  \frac{1}{T} | \mathcal{T}_{\delta, | \Psi (0) \rangle}^{[0, T]} | \rightarrow 0
\end{equation}
in the thermodynamic limit $L\to\infty$.
\end{theorem} 

The theorem does not assume Gaussianity or translation invariance of the initial state. In particular, taking $g(L)=O(\ln L)$, we find that the equilibration time exhibits sublinear scaling: $t_{\mathrm{eq}}=O(L^{\alpha-1/2}\ln L)=o(L)$ for $\alpha<3/2$ in the thermodynamic limit $L\to\infty$, indicating that long-range hopping accelerates equilibration.

\subsection{Density fluctuation and energy-shell counting}
We first perform the Fourier transformation to the fermionic operator
\begin{equation}
  a_{\beta}^{\dagger} = \frac{1}{\sqrt{L}} \sum_{x \in \Lambda_L} e^{2 \pi i
  \beta x / L} c_x^{\dagger},
\end{equation}
where $\beta \in \mathbb{Z}/L\mathbb{Z}$ and $- L / 2 < \beta \leqslant L / 2$. Then the
Hamiltonian \eqref{eq:Heq} can be rewritten as
\begin{equation}
  H = \sum_{\beta} E_{\beta} a_{\beta}^{\dagger} a_{\beta},
\end{equation}
where
\begin{equation}
  E_{\beta} = \sum_{r = 1}^{L - 1} J_L (r) e^{- i 2 \pi \beta r / L} =
  E_{\beta}^{(0)} + \frac{1}{t_L} \cos \theta_{\beta} - \frac{J_0}{L^{\alpha}}, \label{eq:dispersion}
\end{equation}
with
\begin{eqnarray}
  E_{\beta}^{(0)} & := & 2 J_0 \sum^{L / 2 - 1}_{r = 1} \frac{\cos (2 \pi
  \beta r / L)}{r^{\alpha}} + \frac{J_0}{(L / 2)^{\alpha}} (- 1)^{\beta}, \\
  \theta_{\beta} & : = & \frac{\pi}{4} - \frac{\pi \beta^2}{L}, \\
  t_L & := & L^{\alpha - 1 / 2} / J_0 . 
\end{eqnarray}
The oscillatory contribution in Eq. \eqref{eq:dispersion} arises from the complete quadratic Gauss sum
\begin{equation}
    \sum_{r=0}^{L-1}e^{i\pi r^2/L-2\pi i\beta r/L}=\sqrt{L}e^{i\pi/4-i\pi\beta^2/L},
\end{equation}
and the subtraction $-J_0/L^{\alpha}$ removes the $r=0$ term. By the definition of the nonequilibrium subspace,
\begin{equation}
  \langle \Psi (t) | \sum_c (\Delta \rho_c)^2 | \Psi (t) \rangle \geqslant
  \langle \Psi (t) | P_{\text{neq}} \sum_c (\Delta \rho_c)^2 | \Psi (t)
  \rangle \geqslant (\varepsilon \bar{\rho})^2 \langle \Psi (t) |
  P_{\text{neq}} | \Psi (t) \rangle,
\end{equation}
we have
\begin{equation}
  \langle \Psi (t) | P_{\text{neq}} | \Psi (t) \rangle \leqslant
  \frac{1}{(\varepsilon \bar{\rho})^2} \langle \Psi (t) | \sum_c (\Delta
  \rho_c)^2 | \Psi (t) \rangle = \frac{1}{(\varepsilon \bar{\rho})^2} \langle
  \Psi (0) | \sum_c (\Delta \rho_{c, H} (t))^2 | \Psi (0) \rangle,
  \label{eq:Pnequ}
\end{equation}
where we introduce the Heisenberg picture
\begin{equation}
  \Delta \rho_{c, H}  (t) := e^{i H t} \Delta \rho_c e^{- i H t} .
\end{equation}
Thus, the analysis of $t$ satisfying $\langle \Psi (t) | P_{\text{neq}} | \Psi
(t) \rangle>\delta$ is reduced to analyzing the operator $\Delta \rho_{c, H} (t)$.
Following the proof of Proposition 2 in Ref. \cite{HaraKoike2026}, we obtain
\begin{equation}
  \Delta \rho_{c, H}  (t) = \frac{1}{L} \sum_{m \neq 0} \tilde{w}_c (m)
  \sum_{\beta} e^{- i t \tilde{E}_{\beta, m}} a_{\beta + m}^{\dagger}
  a_{\beta} = \frac{1}{L} \sum_{m \neq 0} \tilde{w}_c (m) F (m),
  \label{eq:rhocH}
\end{equation}
where
\begin{equation}\label{eq:wc}
  \tilde{w}_c (m) := \frac{1}{l_c} \sum_{x \in B (c)} e^{- 2 \pi i m x /
  L} = e^{-2\pi imc/L}\frac{\sin (\pi l_c m / L)}{l_c \sin (\pi m / L)},
\end{equation}
for $m \neq 0$ and
\begin{equation}
  F (m) := \sum_{\beta} e^{- i t \tilde{E}_{\beta, m}} a_{\beta +
  m}^{\dagger} a_{\beta},
\end{equation}
with $\tilde{E}_{\beta, m} := E_{\beta} - E_{\beta + m}$. One can find
that
\begin{equation}
  | \tilde{w}_c (m) | \leqslant \min \left( 1, \frac{1}{l_c | \sin (\pi m / L)
  |} \right) \leqslant \min \left( 1, \frac{n}{2 | m |} \right),
  \label{eq:upperw}
\end{equation}
where $0 < |m | \leqslant L / 2 $. Next, we consider the normalized average of an operator-valued function $X (t)$ \cite{HaraKoike2026}
\begin{equation}
  [X]_{\tau} = \int_{- \infty}^{\infty} d t f_{\tau} (t) X (t) = \int_{-
  \infty}^{\infty} d t \frac{1}{\pi \tau} \left( \frac{\sin (t / \tau)}{t /
  \tau} \right)^2 X (t).
\end{equation}
The kernel is nonnegative and integrates to one. Applying this average to the square of Eq. \eqref{eq:rhocH} gives
\begin{equation}
  [(\Delta \rho_{c, H}  (t))^2]_{\tau} = \frac{\tau}{2} \int_{-
  \infty}^{\infty} d E G_c (E) G_c^{\dagger} (E), \label{eq:rho2}
\end{equation}
with
\begin{equation}
  G_c (E) = \frac{1}{L} \sum_{m \neq 0} \tilde{w}_c (m) \sum_{\beta \in
  \Omega_m (E)} a_{\beta + m}^{\dagger} a_{\beta},
\end{equation}
where the energy shell $\Omega_m (E)$ is
\begin{equation}
  \Omega_m (E) := \left\{ \beta : | E - \tilde{E}_{\beta, m} | <
  \frac{1}{\tau} \right\} .
\end{equation}
Since the operator norm of the particle number operator of fermions is $\| n_x
\|_{\text{op}} \leqslant 1$, the Cauchy-Schwarz inequality gives
\begin{equation}
  \| [(\Delta \rho_{c, H}  (t))^2]_{\tau} \|_{\text{op}} \leqslant \left(
  \frac{1}{L} \sum_{m \neq 0} | \tilde{w}_c (m) | \sqrt{I_m} \right)^2,
  \label{eq:rhocHt}
\end{equation}
where
\begin{equation}
  I_m := \frac{\tau}{2} \int_{- \infty}^{\infty} d E | \Omega_m (E) |^2 .
  \label{eq:Im}
\end{equation}
Here we use the equality
\begin{equation}
  [e^{- i t (E' - E'') }]_{\tau} = \frac{\tau}{2} \int_{- \infty}^{\infty} d E
  I \left[ | E - E' | \leqslant \frac{1}{\tau} \right] I \left[ | E - E'' |
  \leqslant \frac{1}{\tau} \right],
\end{equation}
and $I [x \leqslant a] = \theta (a - x)$ is the indicator function. The remaining task is to bound the number of momenta in each energy shell $| \Omega_m (E) |$. The dispersion relation in Eq. \eqref{eq:dispersion} implies that
\begin{equation}
  t_L \tilde{E}_{\beta, m} = \cos \theta_{\beta} - \cos \theta_{\beta + m} +
  t_L (E_{\beta}^{(0)} - E^{(0)}_{\beta + m}) .
\end{equation}
Define
\begin{equation}
  F (z) = \frac{1}{(2 \pi)^2} \int_0^{2 \pi} \int_0^{2 \pi} I [\cos u - \cos v
  \leqslant z] d u d v,
\end{equation}
which satisfies
\begin{equation}
  \sup_z \{ F (z + \delta) - F (z - \delta) \} \leqslant C \sqrt{\delta},
  \label{eq:Fdelta}
\end{equation}
for an absolute constant $C>0$ and $0 \leqslant \delta \leqslant 1$. Next, we define the empirical distribution function
\begin{equation}
  F_{m, L} (z) = \frac{1}{L} | \{ \beta : t_L \tilde{E}_{\beta, m} \leqslant z
  \} | = \frac{1}{L} | \{ \beta : \cos \theta_{\beta} - \cos \theta_{\beta +
  m} + t_L (E_{\beta}^{(0)} - E^{(0)}_{\beta + m}) \leqslant z \} | ,
  \label{eq:FmLz}
\end{equation}
and its uniform deviation from $F(z)$
\begin{equation}
  e_{m, L} := \sup_z | F_{m, L} (z) - F (z) | . \label{eq:emL}
\end{equation}
Indeed, the function $F (z)$ can be understood as the continuous version of
$F_{m, L} (z)$ if we neglect $E_{\beta}^{(0)} - E^{(0)}_{\beta + m}$ and average over $m$. For a sufficiently large $L$ and $0<|m|\leqslant M\leqslant \ln L$, we will prove that
\begin{equation}
  e_{m, L} \leqslant \tilde{C}_{\alpha} \left( \frac{| m | \ln (2 L)}{L^{3 / 2
  - \alpha}} \right)^{1 / 3}, \label{eq:emLth}
\end{equation}
which vanishes in the limit $L \rightarrow \infty$. This bound ensures that $F_{m, L} (z) \rightarrow F (z)$ for every $z$ in the thermodynamic limit. 
\subsection{Distribution of rescaled energy differences}
To show Eq. \eqref{eq:emLth}, we first estimate $| t_L (E_{\beta}^{(0)} - E^{(0)}_{\beta + m})
|$:
\begin{equation}
  | t_L (E_{\beta}^{(0)} - E^{(0)}_{\beta + m}) | = 2 J_0 t_L \left| \sum^{L /
  2 - 1}_{r = 1} \frac{[\cos (2 \pi \beta r / L) - \cos (2 \pi (\beta + m) r /
  L)]}{r^{\alpha}} + \frac{1}{2 (L / 2)^{\alpha}} (- 1)^{\beta} [1 - (- 1)^m]
  \right| .
\end{equation}
The second term on the right-hand side of this equation can be upper bounded by
\begin{equation}\label{eq:E01}
  \left| \frac{J_0 t_L}{(L / 2)^{\alpha}} (- 1)^{\beta} [1 - (- 1)^m] \right|
  \leqslant 2^{\alpha + 1} L^{- 1 / 2} \rightarrow 0.
\end{equation}
The remaining sum gives the result
\begin{equation}
  \left| \sum^{L / 2 - 1}_{r = 1} \frac{[\cos (2 \pi \beta r / L) - \cos (2
  \pi (\beta + m) r / L)]}{r^{\alpha}} \right| = \left| \sum^{L / 2 - 1}_{r =
  1} \frac{[\cos (k_{\beta} r) - \cos (k_{\beta + m} r)]}{r^{\alpha}} \right|,
\end{equation}
where $k_{\beta} := 2 \pi \beta / L$. For $0<k\leqslant\pi$, we have
\begin{equation}
    | d
\mathcal{E}_{\beta}^{(0)} / d k |=2J_0\left|\sum^{L / 2 - 1}_{r = 1}r^{1-\alpha}\sin(kr)\right|\leqslant C'_{\alpha}k^{\alpha-2},
\end{equation}
where $\mathcal{E}_{\beta}^{(0)}:=2J_0\sum_{r=1}^{L/2-1}\cos(kr)/r^{\alpha}$. Thus,
\begin{equation}
    \int_0^{2\pi}\left|\frac{d\mathcal{E}^{(0)}_{\beta}}{dk}\right|dk\leqslant C_{\alpha}'\frac{\pi^{\alpha-1}}{\alpha-1}=:C_{\alpha}.
\end{equation}
Each nearest-neighbor difference on the momentum grid is bounded by the variation over the corresponding grid interval. Summing these bounds and telescoping over $|m|$ grid steps yields
\begin{equation}
  \frac{1}{L}\sum_{\beta=0}^{L-1}2J_0\left| \sum^{L / 2 - 1}_{r = 1} \frac{[\cos (k_{\beta} r) - \cos (k_{\beta +
  m} r)]}{r^{\alpha}} \right| \leqslant C_{\alpha}  \frac{| m |}{ L} .
\end{equation}
Therefore,
\begin{equation}
   \frac{1}{L}\sum_{\beta=0}^{L-1}\left| t_L(E_{\beta}^{(0)}-E_{\beta+m}^{(0)}) \right| \leqslant C_{\alpha} |
  m | L^{\alpha - 3 / 2} . \label{eq:E0}
\end{equation}
Due to the fact that $1 < \alpha < 3 / 2$, we apply Markov's inequality and obtain
\begin{equation}
   \frac{1}{L} \times | \{ \beta : | t_L
  (E_{\beta}^{(0)} - E^{(0)}_{\beta + m}) | > a \} |\leqslant\frac{C_{\alpha}|m|}{aL^{3/2-\alpha}} \to 0,
\end{equation}
for every $a > 0$ and $m \neq 0$ in the limit $L\to\infty$. There exists a subtlety that a large $m\sim L$ leads to the divergence of the right-hand side of Eq. \eqref{eq:E0}.We here only
restrict the discussion to small $m$ satisfying $0 < |m | \leqslant M
$, where $M$ is the cutoff which diverges in the limit $L
\rightarrow \infty$. Later we will take $M = O (\ln L)$ to prohibit the
divergence of the right-hand side of Eq. \eqref{eq:E0}. 

Separate the oscillatory part of the empirical distribution by defining
\begin{equation}
  F_{m, L}^{\text{osc}} (z) := \frac{1}{L} | \{ \beta : \cos
  \theta_{\beta} - \cos \theta_{\beta + m} \leqslant z \} |,
\end{equation}
with the error distance
\begin{equation}
  e_{m, L}^{\text{osc}} := \sup_z | F_{m, L}^{\text{osc}} (z) - F (z) | .
\end{equation}
We first evaluate $e_{m, L}^{\text{osc}}$. Let $D_{m, L}$ denote the
rectangular discrepancy of the normalized phase-pair points
$\bm{x}_{\beta} := (\theta_{\beta} / 2 \pi, \theta_{\beta + m} / 2
\pi)$ mod 1, which takes the form of
\begin{equation}
  D_{m, L} = \sup_R \left| \frac{1}{L} \sum_{\beta = 0}^{L - 1} \bm{1}_R
  (\bm{x}_{\beta}) - | R | \right|, \label{eq:DmLR}
\end{equation}
where $R = [a, b) \times [c, d)$ is a subset of $[0, 1)^2$, $| R | = (b - a)
(d - c)$ is the area of $R$, and $\bm{1}_R (\bm{x}_{\beta})$ is the indicator function taking the value of 1 if $\bm{x}_{\beta} \in
R$ and 0 otherwise. This function $D_{m, L}$ describes how uniformly the vectors $\bm{x}_{\beta}$ are distributed in the square $[0, 1]^2$ and therefore measures the deviation of the empirical phase-pair distribution from the uniform area measure. Following
the two-dimensional Erdos-Turan-Koksma inequality \cite{Stefan2019}, the function $D_{m,L}$ can be estimated as
\begin{equation}
  D_{m, L} \leqslant C_{\text{ETK}} \left[ \frac{1}{H} + \sum_{| h |, | j |
  \leqslant H ; (h, j) \neq (0, 0)} \frac{| A_{m, L} (h, j) |}{\max (1, | h |)
  \max (1, | j |)} \right], \label{eq:DmL}
\end{equation}
for any integer $H \geqslant 1$, where
\begin{equation}
  A_{m, L} (h, j) = \frac{1}{L} \sum_{\beta = 0}^{L - 1} e^{i (h
  \theta_{\beta} + j \theta_{\beta + m})} .
\end{equation}
Here, the integer $H$ is a cutoff. To evaluate the value of this upper bound, we introduce
\begin{equation}\label{eq:SL}
  S_L (a, b) = \sum_{\beta = 0}^{L - 1} \exp \left[ - \frac{i \pi}{L} (a
  \beta^2 + 2 b \beta) \right] .
\end{equation}
An explicit calculation shows that
\begin{equation}
  A_{m, L} (h, j) = \frac{e^{i (h + j) \pi / 4 - i \pi j m^2 / L}}{L} S_L (h +
  j, j m) .
\end{equation}
Because $L$ is even, the summand in Eq. \eqref{eq:SL} is periodic in $\beta$ modulo $L$. Expanding the absolute square and summing over one index gives 
\begin{equation}
  | S_L (a, b) |^2 = L \sum_{0 \leqslant s < L, a s \equiv 0 (\text{mod} L)}
  e^{- i \pi (a s^2 + 2 b s) / L} .
\end{equation}
Hence,
\begin{equation}
  | S_L (a, b) | \leqslant \sqrt{L \gcd (| a |, L)} .
\end{equation}
For $a = 0$, the sum vanishes unless $L | b $, in which case $| S_L
(0, b) | = L$. In particular,
\begin{equation}
  | A_{m, L} (h, j) | \leqslant \frac{| S_L (h + j, j m) |}{L} .
\end{equation}
For $|h|,|j|\leqslant H$ and $h+j\neq0$, we have
\begin{equation}
  | A_{m, L} (h, j) | \leqslant \frac{\sqrt{\gcd (| h + j |, L)}}{\sqrt{L}}
  \leqslant \sqrt{\frac{2 H}{L}} .
\end{equation}
For $h+j=0$ and $(h,j)\neq(0,0)$, we have $j\neq0$ and $S_L(0,jm)=L$ if $L\mid jm$, and zero otherwise. When we take $0<|j|\leqslant H=\lfloor L^{1 / 3} \rfloor$, we have $0<|jm|\leqslant H^2<L$ and consequently, $L\nmid|jm|$. Hence, $| A_{m, L} (h, j) |\leqslant \sqrt{2H/L}$ holds for all $|h|,|j|\leqslant H$ with $(h,j)\neq(0,0)$.

Next, the sum in Eq. \eqref{eq:DmL} is bounded by
\begin{eqnarray}
  \sum_{| h |, | j | \leqslant H ; (h, j) \neq (0, 0)} \frac{| A_{m, L} (h, j)
  |}{\max (1, | h |) \max (1, | j |)} & \leqslant & \sqrt{\frac{2 H}{L}}
  \sum_{| h |, | j | \leqslant H ; (h, j) \neq (0, 0)} \frac{1}{\max (1, | h
  |) \max (1, | j |)} \nonumber\\
  & = & \sqrt{\frac{2 H}{L}} \left[ \left( \sum_{h = - H}^H \frac{1}{\max (1,
  | h |)} \right)^2 - 1 \right] \nonumber\\
  & = & \sqrt{\frac{2 H}{L}} \left[ \left( 1 + 2 \sum_{h = 1}^H \frac{1}{h}
  \right)^2 - 1 \right] 
\end{eqnarray}
For every integer $H\geqslant1$, the harmonic sum gives
\begin{equation}
  \sum_{h = 1}^H \frac{1}{h} \leqslant 1 + \ln H .
\end{equation}
Hence,
\begin{equation}
  D_{m, L} \leqslant C'_{\text{ETK}} \left[ \frac{1}{H} + \sqrt{\frac{H}{L}}
  (1 + \ln H)^2 \right] .
\end{equation}
If we take $H = \lfloor L^{1 / 3} \rfloor$, we have the bound
\begin{equation}
  D_{m, L} \leqslant \tilde{C}_{\text{ETK}} L^{- 1 / 3} [\ln (2 L)]^2,
  \label{eq:DmLf}
\end{equation}
which vanishes in the limit $L \rightarrow \infty$.

Then we relate $D_{m, L}$ to $e_{m, L}^{\text{osc}}$. Define the empirical
counting measure
\begin{equation}
  \nu_{m, L} (A) = \frac{1}{L} | \{ \beta : (\theta_{\beta}, \theta_{\beta +
  m}) \text{mod} 2 \pi \in A \} |,
\end{equation}
and the normalized area measure
\begin{equation}
    \lambda (A) = \frac{| A |}{(2 \pi)^2} .
\end{equation}
Compared with the definition of $D_{m, L}$ in Eq. \eqref{eq:DmLR}, we find
that
\begin{equation}
  D_{m, L} = \sup_R | \nu_{m, L} (R) - \lambda (R) |,
\end{equation}
where $R$ now ranges over axis-aligned rectangles in $[0, 2 \pi)^2$. On the other hand, the error $e_{m, L}^{\text{osc}}$ concerns
\begin{equation}
  e_{m, L}^{\text{osc}} = \sup_z | \nu_{m, L} (A_z) - \lambda (A_z) |,
\end{equation}
with
\begin{equation}
  A_z := \{ (u, v) \in [0, 2 \pi)^2 : \cos u - \cos v \leqslant z \} .
\end{equation}
Notice that $A_z$ is not a rectangular region. By partitioning the phase domain into monotonicity regions, we construct upper and lower approximations to \(A_z\) using at most \(4K\) rectangles, with an area difference bounded by \(1/K\)~\cite{Heiner_1975,Harald1992}:
\begin{equation}
  | \nu_{m, L} (A_z) - \lambda (A_z) | \leqslant 4 K D_{m, L} + \frac{1}{K},
\end{equation}
for every $K \geqslant 1$. Therefore,
\begin{equation}
  e_{m, L}^{\text{osc}} \leqslant 4 K D_{m, L} + \frac{1}{K} .
\end{equation}
Here we choose $K = \lceil D_{m, L}^{- 1 / 2} \rceil$ for $D_{m,L}>0$. Since $0<D_{m,L}\leqslant1$, we have
\begin{equation}
  e_{m, L}^{\text{osc}} \leqslant 4 K D_{m, L} + \frac{1}{K} \leqslant 4
  \left( \sqrt{D_{m, L}} + D_{m, L} \right) + \sqrt{D_{m, L}} \leqslant 9
  \sqrt{D_{m, L}} . \label{eq:eosc}
\end{equation}
Substituting Eq. \eqref{eq:DmLf} into Eq. \eqref{eq:eosc} yields
\begin{equation}
  e_{m, L}^{\text{osc}} \leqslant C_{\text{osc}} L^{- 1 / 6} \ln (2 L) .
\end{equation}

Then we return to estimate $e_{m, L}$ in Eq. \eqref{eq:emL}. We first define
\begin{equation}
  e_{m, L}^{(0)} := \sup_z | F_{m, L} (z) - F_{m, L}^{\text{osc}} (z) |
\end{equation}
and denote
\begin{equation}
  b_{m, L} := \frac{1}{L} \sum_{\beta = 0}^{L - 1} | t_L (E_{\beta}^{(0)}
  - E_{\beta + m}^{(0)}) |.\label{eq:b/L}
\end{equation}
For every $h>0$, Eq. \eqref{eq:b/L} implies that the fraction of $\beta$ in $[0, L - 1]$ satisfying $| t_L
(E_{\beta}^{(0)} - E_{\beta + m}^{(0)}) | > h$ is at most $b_{m, L} / h$. Recalling the definition of $F_{m, L} (z)$ in Eq. \eqref{eq:FmLz}, we obtain
\begin{equation}\label{eq: Fud}
  F_{m, L}^{\text{osc}} (z - h) - \frac{b_{m, L}}{h} \leqslant F_{m, L} (z)
  \leqslant F_{m, L}^{\text{osc}} (z + h) + \frac{b_{m, L}}{h} .
\end{equation}
Equation \eqref{eq:Fdelta} applied with interval midpoint $z\pm h/2$ and half-width $h/2$ bounds the increments of $F(z)$ by $C\sqrt{h/2}$. Since $F_{m,L}^{\text{osc}}(z)$ differs from $F(z)$ by at most $e_{m,L}^{\text{osc}}$, Eq. \eqref{eq: Fud} can be rewritten as
\begin{equation}
  F_{m, L}^{\text{osc}} (z) - C \sqrt{h / 2} -2e_{m,L}^{\text{osc}}- \frac{b_{m, L}}{h} \leqslant
  F_{m, L} (z) \leqslant F_{m, L}^{\text{osc}} (z) + C \sqrt{h / 2} +2e_{m,L}^{\text{osc}}+
  \frac{b_{m, L}}{h} .
\end{equation}
Therefore,
\begin{equation}
  e_{m, L}^{(0)} \leqslant C \sqrt{h} +2e_{m,L}^{\text{osc}}+ \frac{b_{m, L}}{h} .
\end{equation}
Taking $h = b_{m, L}^{2 / 3}$ gives
\begin{equation}
  e_{m, L}^{(0)} \leqslant (C + 1) b_{m, L}^{1 / 3}+2e_{m,L}^{\text{osc}} .
\end{equation}
Together with Eq. \eqref{eq:eosc}, we can estimate $e_{m,L}$ as
\begin{equation}
  e_{m, L} \leqslant e_{m, L}^{(0)} + e_{m, L}^{\text{osc}} \leqslant (C + 1)
  b_{m, L}^{1 / 3} + 3C_{\text{osc}} L^{- 1 / 6} \ln (2 L) .
\end{equation}
Substituting Eqs. \eqref{eq:E01} and \eqref{eq:E0} into Eq. \eqref{eq:b/L} yields
\begin{equation}
  b_{m,L}=\frac{1}{L} \sum_{\beta = 0}^{L - 1} | t_L (E_{\beta}^{(0)} - E_{\beta +
  m}^{(0)}) | \leqslant C_{\alpha} | m | L^{\alpha - 3 / 2},
\end{equation}
we obtain
\begin{equation}
  e_{m, L} \leqslant C_{\alpha}^{1/3} (C + 1) | m |^{1 / 3} L^{(\alpha - 3 / 2) / 3}
  + C_{\text{osc}} L^{- 1 / 6} \ln (2 L) . \label{eq:emLfinal}
\end{equation}
Since $1 < \alpha < 3 / 2$, we have $0 < (3 / 2 - \alpha) / 3 < 1 /
6$. Thus, the first term on the right-hand side of Eq. \eqref{eq:emLfinal}
dominates in the limit $L \rightarrow \infty$. For a sufficiently large $L$,
$e_{m, L}$ can be bounded as
\begin{equation}
  e_{m, L} \leqslant \tilde{C}_{\alpha} \left( \frac{| m | \ln (2 L)}{L^{3 / 2
  - \alpha}} \right)^{1 / 3},
\end{equation}
by choosing $\tilde{C}_{\alpha} = 2 C_{\alpha}^{1/3} (C + 1)$, which completes the proof of Eq. \eqref{eq:emLth}.

\subsection{Fraction of nonequilibrium times}
For an interval $[a, b]$, the definition of $e_{m,L}$ implies
\begin{equation}
  F_{m, L} (b) - F_{m, L} (a) \leqslant F (b) - F (a) + 2 e_{m, L} .
\end{equation}
In the concerned energy shell: $| t_L E - t_L\tilde{E}_{\beta,m} | \leqslant t_L / \tau$, we choose $a = t_L E - t_L / \tau$ and $b = t_L E + t_L / \tau$. Thus, by using the inequality \eqref{eq:Fdelta}, we obtain
\begin{equation}
  \sup_E \frac{| \Omega_m (E) |}{L} \leqslant C \sqrt{\frac{t_L}{\tau}} + 2
  e_{m, L}, \label{eq:Omega/L}
\end{equation}
where we assume that $\tau \geqslant t_L$. Each momentum $\beta$ provides an interval of a window of length $2 / \tau$, and therefore
\begin{equation}
  \int_{- \infty}^{\infty} d E | \Omega_m (E) | = \frac{2 L}{\tau} .
  \label{eq:iOmega}
\end{equation}
Substituting Eqs. \eqref{eq:Omega/L} and \eqref{eq:iOmega} into Eq.
\eqref{eq:Im} yields
\begin{equation}
  I_m \leqslant L \sup_E | \Omega_m (E) | \leqslant L^2 \left( C
  \sqrt{\frac{t_L}{\tau}} + 2 e_{m, L} \right) .
\end{equation}

Further, we return to analyze the decomposition of Eq. \eqref{eq:rhocH}. To use the estimate only in the range of controlled $m$, we split the density fluctuation into low- and high-momentum parts: $0 < | m | \leqslant M$
and $M < | m | \leqslant L / 2$. Thus, we can reorganize $\Delta \rho_c$ as
\begin{equation}
  \Delta \rho_c = \Delta \rho_c^{(M)} + R_{c, M},
\end{equation}
where
\begin{eqnarray}
  \Delta \rho_c^{(M)} & = & \frac{1}{L} \sum_{m : 0 < | m | \leqslant M}
  \sum_x \tilde{w}_c (m) e^{2 \pi i m x / L} n_x, \\
  R_{c, M} & = & \frac{1}{L} \sum_{m : M < | m | \leqslant L / 2} \sum_x
  \tilde{w}_c (m) e^{2 \pi i m x / L} n_x =: \frac{1}{L} \sum_x g_{c,
  M} (x) n_x, 
\end{eqnarray}
with
\begin{equation}
  g_{c, M} (x) := \sum_{m :| m |>M} \tilde{w}_c (m)
  e^{2 \pi i m x / L} .
\end{equation}
Since $\| n_x\|_{\text{op}} \leqslant 1$ for fermions, the Cauchy-Schwarz inequality and the discrete Parseval identity imply
\begin{eqnarray}
  \| R_{c, M} \|_{\text{op}} & \leqslant & \frac{1}{L} \sum_x | g_{c, M} (x) |
  \nonumber\\
  & \leqslant & \left( \frac{1}{L} \sum_x | g_{c, M} (x) |^2 \right)^{1 / 2}
  \nonumber\\
  & = & \left(  \sum_{m :\  | m | >M} | \tilde{w}_c
  (m) |^2 \right)^{1 / 2} \nonumber\\
  & \leqslant & \frac{n}{\sqrt{2 M}}, 
\end{eqnarray}
where we apply Eq. \eqref{eq:upperw} and $\sum_{m : | m | > M} m^{- 2}
\leqslant 2 M^{- 1}$ in the last inequality. Unitary invariance of the operator norm indicates that
\begin{equation}
  \| R_{c, M, H} (t) \|_{\text{op}}^2 \leqslant \frac{n^2}{2 M} .
\end{equation}
From Eq. (49) of Ref. \cite{HaraKoike2026}, we find that summing Eq. \eqref{eq:wc} gives the bound
\begin{equation}
  \sum_{m : 0 < | m | \leqslant M} | \tilde{w}_c (m) | \leqslant n + 2 + n \ln
  (M) .
\end{equation}
Substituting this equation into Eq. \eqref{eq:rhocHt} and restricting the sum to $0 <| m | \leqslant M$ gives rise to
\begin{eqnarray}
  \| [(\Delta \rho^{(M)}_{c, H}  (t))^2]_{\tau} \|_{\text{op}} & \leqslant &
  \left( \frac{1}{L} \sum_{m:0<|m|\leqslant M} | \tilde{w}_c (m) | \sqrt{I_m} \right)^2
  \nonumber\\
  & \leqslant & (n + 2 + n \ln M)^2 \left( C \sqrt{\frac{t_L}{\tau}} + 2
  e_{M, L} \right), 
\end{eqnarray}
where $e_{M, L} := \max_{0 < |m | \leqslant M }
e_{m, L}$. Applying the inequality $(A + B)^2 \leqslant 2 A^2 + 2 B^2$ for
the Hermitian operators $A$ and $B$, together with
positivity of the averaging kernel, yields
\begin{eqnarray}
  \| [(\Delta \rho_{c, H}  (t))^2]_{\tau} \|_{\text{op}} & \leqslant & 2 \|
  [(\Delta \rho^{(M)}_{c, H}  (t))^2]_{\tau} \|_{\text{op}} + 2 \| R_{c, M, H}
  (t) \|_{\text{op}}^2 \nonumber\\
  & \leqslant & 2 (n + 2 + n \ln M)^2 \left( C \sqrt{\frac{t_L}{\tau}} + 2
  e_{M, L} \right) + \frac{n^2}{M} . \label{eq:rhoch} 
\end{eqnarray}
By taking $\tau = L^{\alpha - 1 / 2} g(L) / J_0$ and $M = \lfloor \min\{\ln L,g(L)\}\rfloor$ with $\lim_{L\to\infty}g(L)=\infty$, the right-hand side of this equation vanishes in the limit $L \rightarrow \infty$.
Denoting
\begin{equation}
  \delta_M := 6 (1 + 2 / n + \ln M)^2 \left( C \sqrt{\frac{t_L}{\tau}}
  + 2 e_{M, L} \right) + \frac{3}{M},
\end{equation}
we substitute Eq. \eqref{eq:rhoch} into Eq. \eqref{eq:Pnequ} and obtain
\begin{equation}
  [\langle \Psi (t) | P_{\text{neq}} | \Psi (t) \rangle]_{\tau} \leqslant
  \frac{1}{(\varepsilon \bar{\rho})^2} \sum_c \| [(\Delta \rho_{c, H} 
  (t))^2]_{\tau} \|_{\text{op}} \leqslant \frac{\delta_M}{3 (\varepsilon
  \bar{\rho})^2} n^3 .
\end{equation}

Finally, we introduce Lemma 4 in Ref. \cite{HaraKoike2026}

\begin{lemma}
Let $X (t)$ be a measurable function satisfying $X (t) > 0$ and $[X (t)]_{\tau} \leqslant B$ with a constant $B > 0$. Then for any positive constant $X_1 > 0$, the fraction of the total time duration with $X(t) > X_1$ in the time interval $[- \tau, \tau]$ is less than $3 B / X_1$.
\end{lemma} 
Taking $X (t) = \langle \Psi (t) | P_{\text{neq}} | \Psi (t)
\rangle$, $X_1 = \sqrt{\delta_M}$, $B = \delta_M n^3 / (3 (\varepsilon
\bar{\rho})^2)$, the measure of the nonequilibrium-time set
$\mathcal{T}_{\sqrt{\delta_M}, | \Psi (0) \rangle}^{[- \tau, \tau]}$ satisfies
\begin{equation}
  \frac{1}{2 \tau} \left| \mathcal{T}_{\sqrt{\delta_M}, | \Psi (0)
  \rangle}^{[- \tau, \tau]} \right| \leqslant \frac{\sqrt{\delta_M}
  n^3}{(\varepsilon \bar{\rho})^2} .
\end{equation}
Since both $n$ and $\varepsilon \bar{\rho}$ are $O (1)$ constants, we take the
thermodynamic limit and conclude that
\begin{equation}
  \frac{1}{T} \left| \mathcal{T}_{\delta, |\Psi(0)\rangle}^{[0, T]} \right|\leqslant\frac{1}{T} \left| \mathcal{T}_{\delta, |\Psi(0)\rangle}^{[-T, T]} \right|\leqslant\frac{2\delta
  n^3}{(\varepsilon \bar{\rho})^2} \rightarrow 0,
\end{equation}
where $\delta := \sqrt{\delta_M} \rightarrow 0$, $T = L^{\alpha - 1 / 2}g(L) / J_0$, and $M = \lfloor \min\{\ln L,g(L)\}\rfloor$. This completes the proof. In particular, choosing $g(L)=\ln L$ yields
\begin{equation}
    T=\frac{L^{\alpha-1/2}\ln L}{J_0}\to\frac{L^{\alpha-1/2}}{J_0}\ll\frac{L}{J_0},
\end{equation}
for $1<\alpha<3/2$ in the limit $L\to\infty$. Thus, long-range hopping can accelerate macroscopic equilibration to a parametrically sub-linear timescale, demonstrating super-ballistically accelerated relaxation.

\end{document}